\documentclass[10pt,aps,pra,amsmath,amssymb,twocolumn,notitlepage,nofootinbib,superscriptaddress]{revtex4-1}
\usepackage{amsmath,graphicx,upgreek,soul,subfigure,float,tabularx,multirow}
\usepackage{wasysym,dcolumn}
\usepackage[colorlinks, linkcolor=blue, citecolor=blue, urlcolor=blue, breaklinks=true]{hyperref}

\def\rmd{\mathrm{d}}
\def\ie{i.e.}
\def\eg{e.g.}

\newcommand{\erefs}[1]{Eqs.~(\ref{#1})}
\newcommand{\fref}[1]{Fig.~\ref{#1}}
\newcommand{\frefs}[1]{Figs.~\ref{#1}}
\newcommand{\tref}[1]{Table~\ref{#1}}

\newcommand{\tp}{\mathrm{T}}

\renewcommand{\vec}[1]{\boldsymbol{#1}}
\newcommand{\mat}[1]{\mathbf{#1}}

\begin{document}

\title{Multipartite optomechanical entanglement from competing nonlinearities}
\author{Andr\'e Xuereb}
\email[Electronic address:\ ]{andre.xuereb@qub.ac.uk}
\affiliation{Centre for Theoretical Atomic, Molecular and Optical Physics, School of Mathematics and Physics, Queen's University Belfast, Belfast BT7\,1NN, United Kingdom}
\author{Marco Barbieri}
\affiliation{Clarendon Laboratory, Department of Physics, University of Oxford, OX1 3PU, United Kingdom}
\author{Mauro Paternostro}
\affiliation{Centre for Theoretical Atomic, Molecular and Optical Physics, School of Mathematics and Physics, Queen's University Belfast, Belfast BT7\,1NN, United Kingdom}
\date{\today}

\makeatletter
\newlength \figurewidth
\setlength \figurewidth {0.45\textwidth}
\makeatother

\begin{abstract}
We investigate the nature of the three-mode interaction inside an optomechanically-active microtoroid with a sizeable $\chi^{(2)}$ coefficient. Experimental techniques are quickly advancing to the point where structures with the necessary properties can be made, and we argue that these provide a natural setting in which to observe rich dynamics leading, for instance, to genuine tripartite steady-state entanglement. We also show that this approach lends itself to a full characterisation of the three-mode state of the system.
\end{abstract}

\maketitle

Over the last several years, the field of optomechanics has witnessed remarkable progress in experimental achievements~\cite{Kippenberg2008,Aspelmeyer2010}, the chief driving factor behind which was the quest to achieve ground-state cooling of a mechanical oscillator. This was achieved, first in by means of cryogenics~\cite{OConnell2010}, then by electromagnetic means in an electromechanical system~\cite{Teufel2011}, and finally in an optomechanical setting~\cite{Chan2011}. Reaching the ground state is a means to an end, for it is only when a mechanical oscillator is close to, or at, the ground state that its true quantum nature shows up unambiguously. This was demonstrated clearly in a recent experiment~\cite{SafaviNaeini2012} that showed the imbalance between the red- and blue-mechanical sidebands in the spectrum, a clear signature that the mechanical oscillator is behaving in a nonclassical way.

\begin{figure}[b]
 \includegraphics[width=\linewidth]{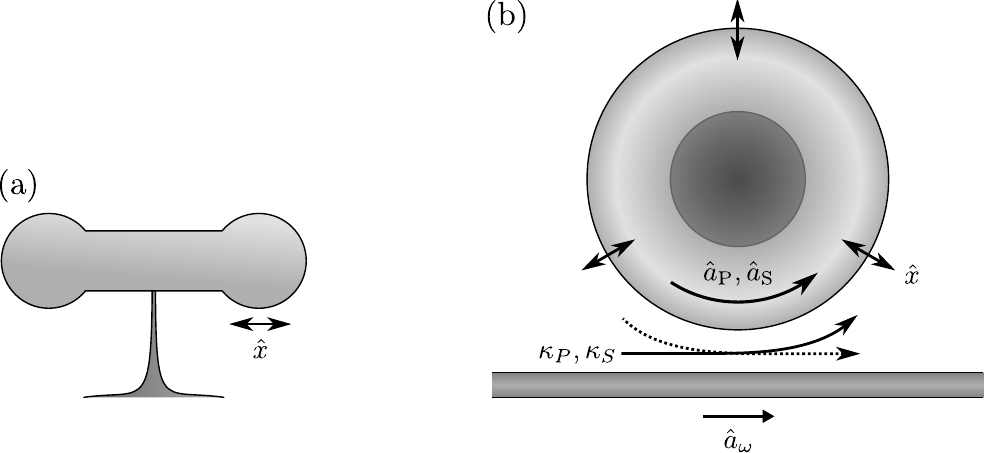}
 \caption{Schematic diagram of the system. (a)~Side view of the toroid. (b)~Top view. We show the coupling of the cavity fields to the field in the waveguide.}
 \label{fig:Model}
\end{figure}

Turning away from fundamental physics, one would like to use quantized mechanical resonators as a \emph{resource}; typical oscillators have decay rates $\kappa_\mathrm{m}$ in the sub-kHz domain, meaning that the decoherence time
\begin{equation}
\tau_\mathrm{d}=\frac{1}{\kappa_\mathrm{m}n_\mathrm{th}}
\end{equation}
can be made large compared to the other timescales of the system by using cryogenic methods to decrease $n_\mathrm{th}=k_\mathrm{B}T_\mathrm{env}\big/(\hbar\omega_\mathrm{m})$, the average number of phonons at an environmental temperature $T_\mathrm{env}$ and at mechanical frequency $\omega_\mathrm{m}$ ($k_\mathrm{B}$ is Boltzmann's constant), as much as possible. Clearly, mechanical oscillators with a large mechanical frequency, say $\omega_\mathrm{m}\gtrsim2\pi\times1$\,MHz, and large mechanical quality factor $Q_\mathrm{m}=\omega_\mathrm{m}\big/(2\kappa_\mathrm{m})$ are at an advantage in this respect. At the same time, one would like the mechanical oscillator to interact strongly with an optical resonator that has a similarly large optical $Q$. It is in this context that optomechanical toroidal structures~\cite{Schliesser2009} appear as ideal optomechanical systems. From a technological point of view, toroidal structures are also ideal in that they minimize the number of moving parts---there are no moving mirrors to align---and can be manufactured monolithically on CMOS-compatible substrates~\cite{Levy2011}, pointing the way towards a possible integration with conventional (opto)electronics in the future.\\
The highest-quality optical modes in toroidal structures are of the `whispering gallery mode' type, with the mechanics of total internal reflection ensuring that losses are minimized. A recent group of experiments~\cite{Furst2010,Levy2011} has recognized this feature as enabling another technology:\ second-harmonic generation (SHG). Indeed, it turns out that toroidal and ring-resonator structures facilitate SHG because the phase matching conditions that are necessary in any nonlinear optics experiment can be met automatically by choosing the right doublet of optical modes~\cite{Furst2010}.\\
It is the purpose of this paper to combine these two ideas. We shall look at the emergence of nonclassical steady-states, \eg, genuinely tripartite entangled states, in the three-mode system formed by the two optical modes---the fundamental and the second harmonic---and the mechanical oscillator. Every pair of these three modes interacts directly, and we shall see that this results in a competition between the two purely optomechanical interactions and the second-harmonic generation process.\par
This paper is structured as follows. Over the next section we shall introduce the full model Hamiltonian, and then proceed to obtain the equations of motion. The usual procedure is used to linearize the dynamics, whereupon we can concentrate exclusively on Gaussian states and present some numerical results. The next section discusses state detection using homodyning techniques, after which we conclude our investigation.

\section{Proposed model}
The model Hamiltonian we use is a combination of the usual optomechanical Hamiltonian and the Hamiltonian description of the SHG process, and describes the system shown schematically in \fref{fig:Model}. We shall label the annihilation operators of the two optical fields $\hat{a}_\mathrm{F}$ (fundamental `F', of frequency $\omega_\mathrm{c}$) and $\hat{a}_\mathrm{S}$ (second harmonic `S', frequency $2\omega_\mathrm{c}$). These two modes are coupled to a continuum of modes, represented by the operators $\hat{a}_\omega$, through decay rates $\kappa_\mathrm{F,S}$ (in practice, $\kappa_\mathrm{F}\approx\kappa_\mathrm{S}$~\cite{Furst2010}), as well as to each other through a second-harmonic interaction frequency $\chi$. The mechanical oscillator `M' is represented through its dimensionless quadratures $\hat{x}$ and $\hat{p}$, and is characterized by the mechanical frequency $\omega_\mathrm{m}$ and decay rate $\kappa_\mathrm{m}$. We allow $\hat{x}$ to couple to the two optical modes through the coupling constants $g_\mathrm{F,S}$; we shall make the simplifying assumption that, since $g_\mathrm{F,S}$ is proportional to the respective optical frequency, $g_\mathrm{S}=2g_\mathrm{F}$. Thus, we can write the Hamiltonian $\hat{H}$ as a sum of four terms
\begin{equation}
\hat{H}=\hat{H}_\mathrm{free}+\hat{H}_\mathrm{diss}+\hat{H}_\mathrm{OM}+\hat{H}_\mathrm{SHG}
\end{equation}
with the free Hamiltonian (we take units such that $\hbar=1$ throughout the paper)
\begin{equation}
\hat{H}_\mathrm{free}{=}\int\rmd\omega\,\omega\,\hat{a}_\omega^\dagger\hat{a}_\omega{+}\sum_{j=\mathrm{F,S}}\mu_j\omega_\mathrm{c}\,\hat{a}_{j}^\dagger\hat{a}_{j}+\frac{\omega_\mathrm{m}}{2}(\hat{x}^2+\hat{p}^2),
\end{equation}
where $\mu_S=2$ and $\mu_F=1$, the dissipation Hamiltonian
\begin{equation}
\hat{H}_\mathrm{diss}=i\sum_{j=\mathrm{F,S}}\sqrt{\frac{\kappa_\mathrm{j}}{\pi}}\int\limits_{\Omega_\mathrm{j}}\rmd\omega\bigl(\hat{a}_\omega^\dagger\hat{a}_{j}-\hat{a}_\omega\hat{a}_{j}^\dagger\bigr)+\hat{H}_\mathrm{\mathrm{M},diss}\,,
\end{equation}
where we leave the mechanical dissipation Hamiltonian undefined, the optomechanical Hamiltonian $\hat{H}_\mathrm{OM}=-\sum_{j=\mathrm{F,S}}g_{j}\hat{a}_{j}^\dagger\hat{a}_{j}\hat{x}$ and, finally, the SHG Hamiltonian~\cite{Walls1995}
\begin{equation}
\hat{H}_\mathrm{SHG}=i\chi\Bigl[(\hat{a}_\mathrm{F}^\dagger)^2\hat{a}_\mathrm{S}-(\hat{a}_\mathrm{F})^2\hat{a}_\mathrm{S}^\dagger\Bigr]\,.
\end{equation}
In the expression for $\hat{H}_\mathrm{diss}$ we defined two frequency ranges $\Omega_\mathrm{F,S}$, which define the bath modes through which the two modes are damped. Given the very large separation in frequency between the two optical modes, we can justify considering $\Omega_\mathrm{F}$ and $\Omega_\mathrm{S}$ as non-overlapping without violating the requirement that these two frequency ranges must be very large compared to $\kappa_\mathrm{F,S}$ that is necessary to ensure Markovian dynamics. If the two frequency ranges were to have a significant overlap, \eg, in the case of two optical modes spaced by a mechanical frequency that exist in different resonators but are coupled to the same bath, then one must be careful to use correctly modified input--output relations and equations of motion. In our case, the standard input--output relations will be held valid. In the following we will obtain the linearized equations of motion for the system described by this Hamiltonian.
\par
\begin{figure*}[t]
 \centering
 \subfigure[\ Reductions, $Q_\mathrm{m}=5970$]{
 \includegraphics[width=0.5\figurewidth]{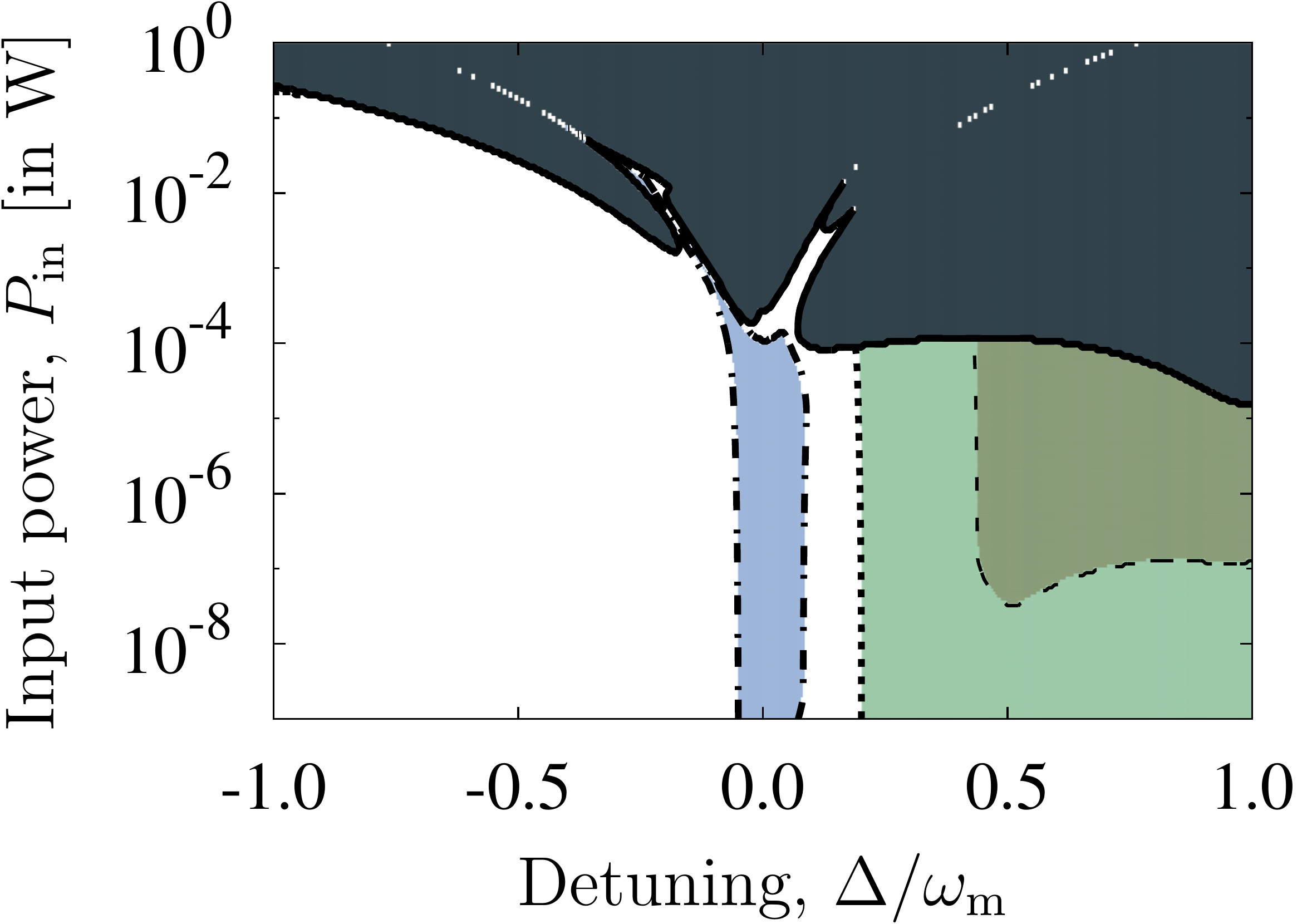}
 }
 \subfigure[\ Bipartitions, $Q_\mathrm{m}=5970$]{
 \includegraphics[width=0.5\figurewidth]{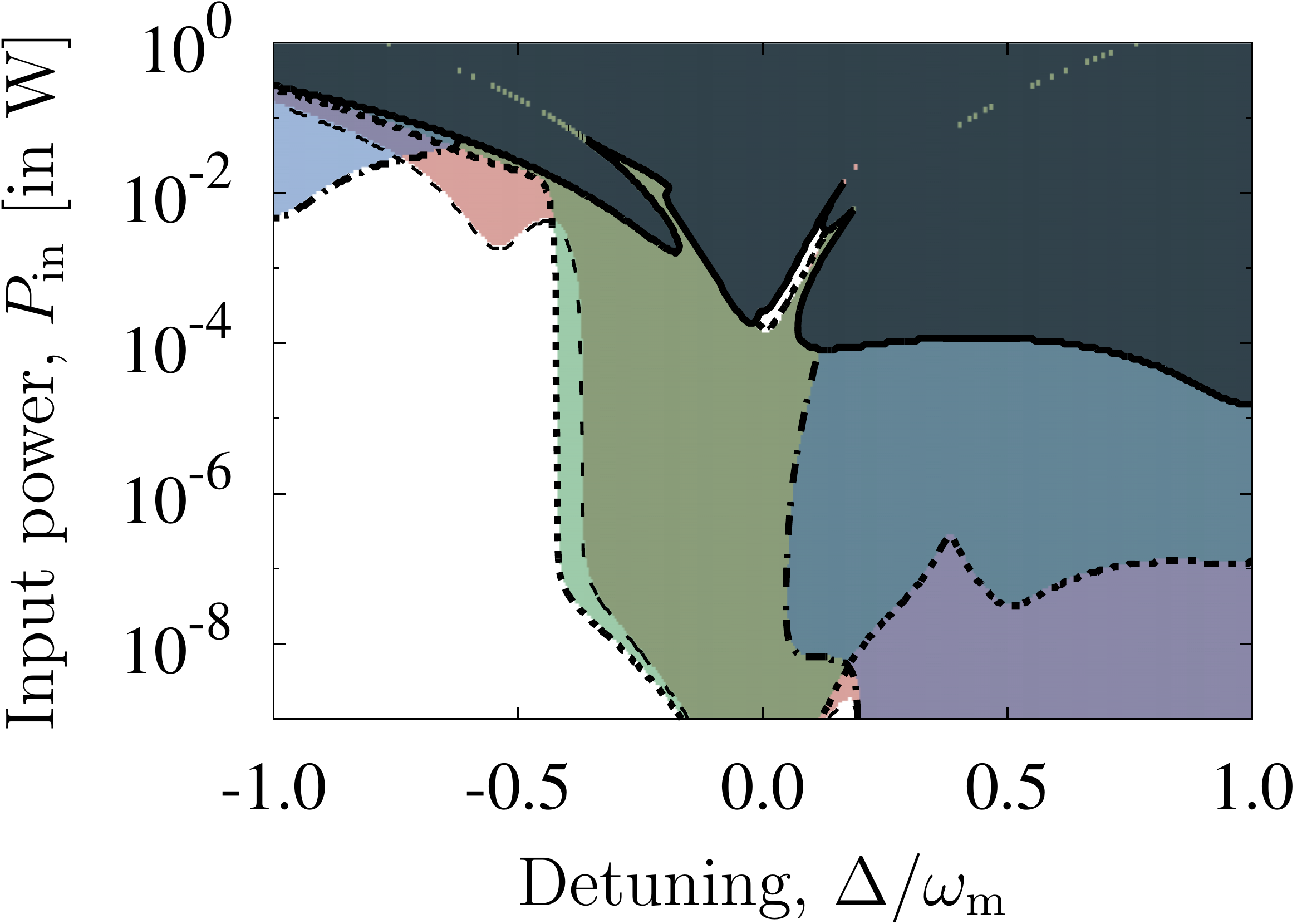}
 }
 \subfigure[\ Reductions, $Q_\mathrm{m}=597000$]{
 \includegraphics[width=.5\figurewidth]{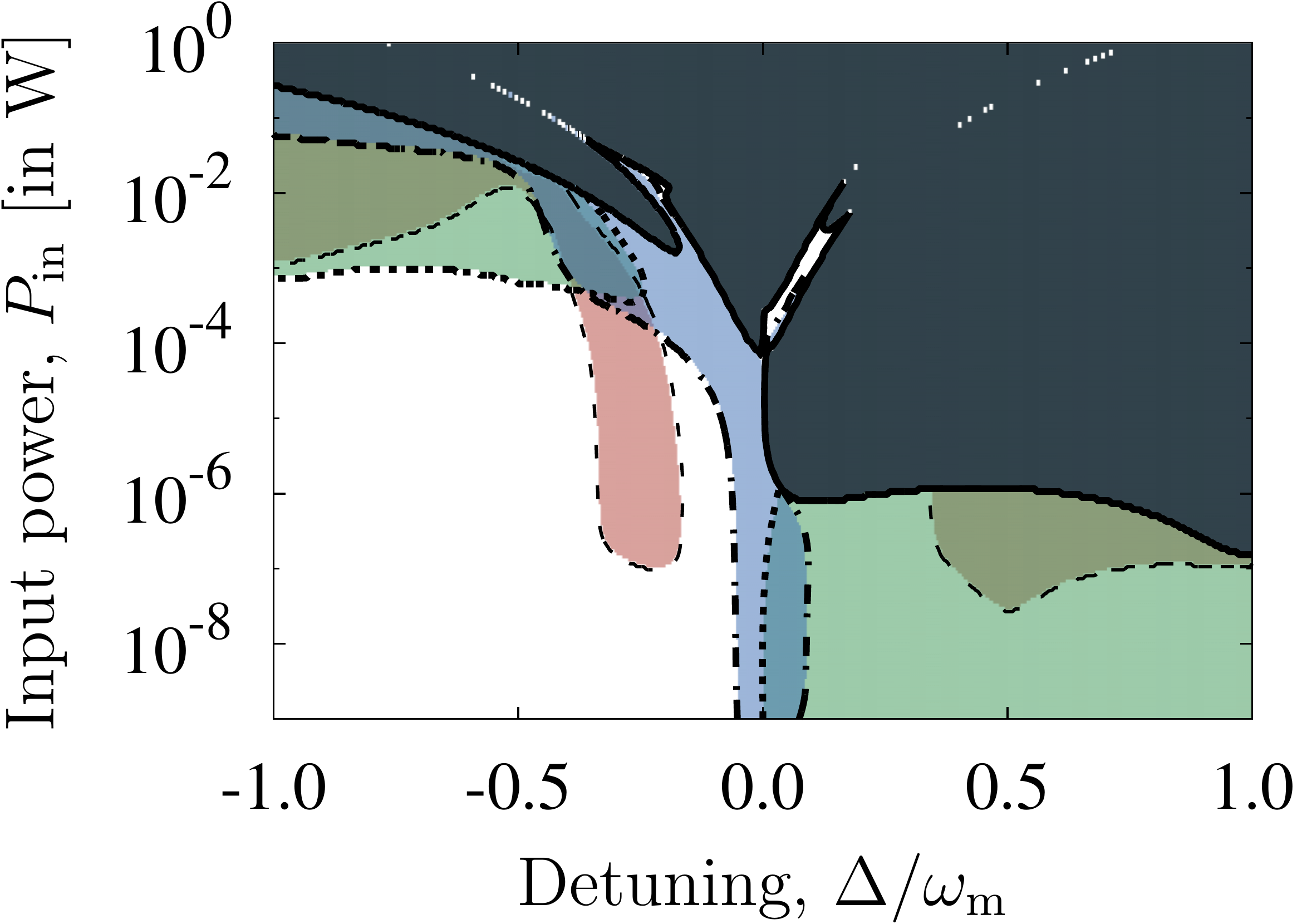}
 }
 \subfigure[\ Bipartitions, $Q_\mathrm{m}=597000$]{
 \includegraphics[width=.5\figurewidth]{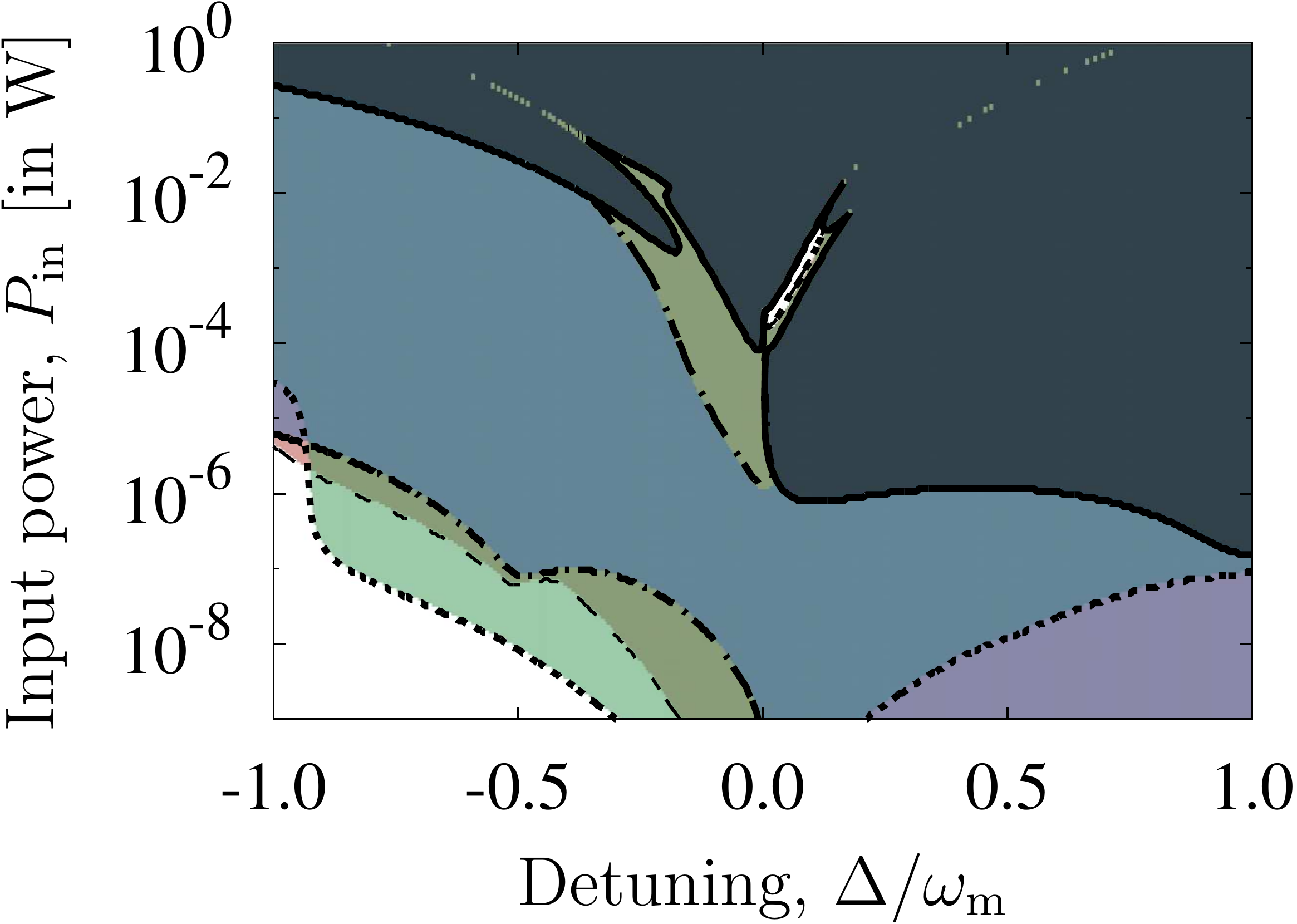}
 }
 \caption{(Color online) Regions of (a)~$1$--$1$-mode (`reduction') entanglement, or (b)~of $1$--$2$-mode (`bipartition') entanglement. In subfigure~(a) the dashed curve encompasses the (red) region where the two-mode reduction consisting of the second-harmonic optical mode (system S) and the mechanical mode M is not separable, the dotted curve encompasses the (green) region where the fundamental optical mode (system F) and the mechanical mode are not separable, and the dashed--dotted curve delineates the (blue) region where the two optical modes are entangled. A similar explanation holds for subfigure~(b), but with the dashed curve corresponding to entanglement between the fundamental optical mode and the $2$-mode system formed by the second-harmonic mode and the mechanical mode. The dotted curve delineates inseparability of the second-harmonic from the $2$-mode system formed by the other two modes, and likewise the dashed--dotted curve bounds the region where the mechanical mode is entangled with the $2$-mode system consisting of the optical modes. The solid curve encloses the region of instability. (c)~and~(d) are similar to (a)~and~(b), respectively, but with a larger mechanical $Q$-factor. Because of the smaller mechanical decay rate, the region of instability starts at significantly lower powers, especially on the blue-detuned side of the figure.}
 \label{fig:EntangledRegions}
\end{figure*}
As a first step we derive the Heisenberg equations of motion for $\hat{a}_{S,F}$, which read:
\begin{equation}
\label{eq:EoM}
\begin{aligned}
\dot{\hat{a}}_\mathrm{F}&=\bigl(i\Delta-\kappa_\mathrm{F}\bigr)\hat{a}_\mathrm{F}+ig_\mathrm{F}\hat{a}_\mathrm{F}\hat{x}+2\chi\hat{a}_\mathrm{F}^\dagger\hat{a}_\mathrm{S}-\sqrt{2\kappa_\mathrm{F}}\hat{a}_\mathrm{F}^\mathrm{in},\\
\dot{\hat{a}}_\mathrm{S}&=\bigl(2i\Delta-\kappa_\mathrm{S}\bigr)\hat{a}_\mathrm{S}+ig_\mathrm{S}\hat{a}_\mathrm{S}\hat{x}-\chi(\hat{a}_\mathrm{F})^2-\sqrt{2\kappa_\mathrm{S}}\,\hat{a}_\mathrm{S}^\mathrm{in}\,,
\end{aligned}
\end{equation}
where $\hat{a}_\mathrm{F(S)}^\mathrm{in}$ is the input field coupled to the fundamental mode (the second harmonic) having zero mean and two-time correlation function
$\langle\hat{a}_\mathrm{F(S)}^{\mathrm{in}\dagger}(t)\hat{a}_\mathrm{F(S)}^\mathrm{in}(t^\prime)\rangle=\delta(t-t^\prime)$ and $\Delta=\omega_\mathrm{F}-\omega_\mathrm{c}$
is the detuning of the driving field from cavity resonance. \erefs{eq:EoM} are written in a frame rotating with the optical modes, \ie, in an interaction picture with respect to the Hamiltonian $\sum_{j=\mathrm{S,F}}\bigl(\int_{\Omega_j}\rmd\omega\,\omega_j\,\hat{a}_\omega^\dagger\hat{a}_\omega+
\omega_j\,\hat{a}_j^\dagger\hat{a}_j\bigr)$. Similarly, for the mechanical mode we have $\dot{\hat{x}}=\omega_\mathrm{m}\hat{p}$ and
\begin{equation}
\dot{\hat{p}}=-\omega_\mathrm{m}\hat{x}-2\kappa_\mathrm{m}\hat{p}-\sqrt{2\kappa_\mathrm{m}}\hat{\xi}+g_\mathrm{F}\hat{a}_\mathrm{F}^\dagger\hat{a}_\mathrm{F}+g_\mathrm{S}\hat{a}_\mathrm{S}^\dagger\hat{a}_\mathrm{S},
\end{equation}
where we used a Brownian-motion--type damping model~\cite{Giovannetti2001}. The self-adjoint Langevin force $\hat{\xi}$ is zero-mean and (assuming the high-temperature limit) delta-correlated as $\langle\hat{\xi}(t)\hat{\xi}(t^\prime)\rangle=(2n_\mathrm{th}+1)\delta(t-t^\prime)$. We now linearize the equations of motion by considering a pumping field of large intensity. Under these conditions, both the field modes of the toroid would be macroscopically populated. We are then allowed to take $\hat{a}_{F,S}=\bar{a}_{F,S}+\delta\hat{a}_{F,S}$, where $\bar{a}_{F,S}=\langle\hat{a}_{F,S}\rangle$ is the (large) mean amplitude of each operator and $\delta\hat{a}$ is its fluctuation around such average. For simplicity of notation, we shall drop the `$\delta$' in the operator fluctuations. We thus write ($j=\mathrm{F,S}$)
\begin{equation}
\dot{\hat{a}}_\mathrm{j}=\bigl[i(\mu_j\Delta+g_\mathrm{j}\bar{x})-\kappa_\mathrm{j}\bigr]\hat{a}_\mathrm{F}+ig_\mathrm{j}\overline{a}_\mathrm{j}\hat{x}+\hat{O}_j-\sqrt{2\kappa_\mathrm{j}}\hat{a}_\mathrm{j}^\mathrm{in}\,
\end{equation}
with $\hat{O}_{\mathrm{F}}=2\chi\bigl(\overline{a}_\mathrm{F}^\ast\hat{a}_\mathrm{S}+\overline{a}_\mathrm{S}\hat{a}_\mathrm{F}^\dagger\bigr)$ and $\hat{O}_{\mathrm{S}}=-2\chi\overline{a}_\mathrm{F}\hat{a}_\mathrm{F}$. Despite the relation linking $g_\mathrm{F}$ and $g_\mathrm{S}$, in what follows we shall continue to use both symbols for clarity. For the mechanical modes, the equation of motion for $\hat{x}$ remains unchanged, while that for momentum becomes
\begin{equation}
\dot{\hat{p}}=-\omega_\mathrm{m}\hat{x}-2\kappa_\mathrm{m}\hat{p}-\sqrt{2\kappa_\mathrm{m}}\hat{\xi}+\sum_{j=\mathrm{F,S}}g_\mathrm{j}\bigl(\overline{a}_\mathrm{j}^\ast\hat{a}_\mathrm{j}+\overline{a}_\mathrm{j}\hat{a}_\mathrm{j}^\dagger\bigr)
\end{equation}
Assuming, for now, the existence of a steady state we find the following equation relating the amplitude of the fundamental mode and that of the input noise
\begin{multline}
\frac{2\chi^2}{(2i\Delta-\kappa_\mathrm{S})}\lvert\overline{a}_\mathrm{F}\rvert^3+\bigl(i\Delta-\kappa_\mathrm{F}\bigr)\lvert\overline{a}_\mathrm{F}\rvert=\sqrt{2\kappa_\mathrm{F}}e^{-i\phi}\overline{a}_\mathrm{in}\,,
\end{multline}
where we have taken $\phi$ is the phase of $\overline{a}_\mathrm{F}$. The input field is assumed to be in a monochromatic coherent state characterized by the (real) amplitude ${\overline{a}_\mathrm{in}}\equiv{\overline{a}_\mathrm{F}^\mathrm{in}}$. For the parameters used throughout this paper, the first term on the left-hand side in the above equation can be safely neglected, whereupon the equation can easily be solved to obtain $\overline{a}_\mathrm{F}$. For the second harmonic, we get
\begin{equation}
{\overline{a}_\mathrm{S}}=\frac{\chi e^{2i\phi}}{2i\Delta-\kappa_\mathrm{S}}\lvert{\overline{a}_\mathrm{F}}\rvert^2,
\end{equation}
while $\bar{x}=\bar{p}=0$. The linearization of the equations of motion makes the dynamics Gaussian so that any initial Gaussian will remain such at any instant of time~\cite{Olivares2011}. We now introduce the quadrature vector
\begin{equation}
\hat{\vec{R}}=\bigl(\hat{x}_\mathrm{F},\hat{p}_\mathrm{F},\hat{x}_\mathrm{S},\hat{p}_\mathrm{S},\hat{x},\hat{p}\bigr)^\tp\,,
\end{equation}
where we have defined the optical quadrature operators $\hat{x}_\mathrm{j}=\frac{1}{\sqrt{2}}(\hat{a}_\mathrm{i}+\hat{a}_\mathrm{i}^\dagger)$ and $\hat{p}_\mathrm{j}=\frac{i}{\sqrt{2}}(\hat{a}^\dag_\mathrm{j}-\hat{a}_\mathrm{j})$ of mode $j=\mathrm{F,S}$. Similar definitions hold for the input-field operators. It is worth bearing in mind that all our operators are functions of time $t$, and we have simply dropped the label $t$ for conciseness of notation. With these definitions, the first moment of $\hat{\vec{R}}$ is zero and any Gaussian state of the system is thus fully characterized by the covariance matrix $\mat{V}=(\langle\hat{\vec{R}}\otimes\hat{\vec{R}}\rangle+\langle\hat{\vec{R}}\otimes\hat{\vec{R}}\rangle^\tp)/2$. The equations of motion derived above can be concisely written as $\dot{\hat{\vec{R}}}=\mat{A}\cdot\hat{\vec{R}}+\hat{\vec{R}}_\mathrm{in}$, with the input noise vector
\begin{equation}
\hat{\vec{R}}_\mathrm{in}{=}(\sqrt{2\kappa_\mathrm{F}}\hat{x}_\mathrm{F}^\mathrm{in},\sqrt{2\kappa_\mathrm{F}}\hat{p}_\mathrm{F}^\mathrm{in},
\sqrt{2\kappa_\mathrm{S}}\hat{x}_\mathrm{S}^\mathrm{in},\sqrt{2\kappa_\mathrm{S}}\hat{p}_\mathrm{S}^\mathrm{in},0,\sqrt{2\kappa_\mathrm{m}}\hat{\xi})^\tp.
\end{equation}
The drift matrix $\mat{A}$ can be explicitly determined and depends on the set of parameters characterizing the dynamics of the three-mode system addressed here. Its expression is too lengthy to be reported here and is thus deferred to the Appendix. A close inspection of the form of $\mat{A}$ reveals that, by assuming $\overline{a}_{\mathrm{in}}\in\mathbb{R}$ and introducing the rescaled parameters $\alpha=g_\mathrm{F}\big/(\sqrt{2}\,\chi)$ and $\beta=\chi{\overline{a}_\mathrm{F}}$, the drift matrix is a universal expression of $g_F/\chi$. Thus, for a fixed value of $\beta$ the nature of the dynamics is determined solely by the ratio of the coupling constants. For small $\alpha$, the interaction is dominated by the SHG process. Conversely, for $\alpha\gg1$, the dynamics resembles closely that of a standard optomechanics model with two fields~\cite{Paternostro2007}. The dynamical equations should be stable in order for a steady state to exist. This is assured if the real part of the spectrum of $\mat{A}$ is negative, in which case the system will tend to a stationary state characterized by the covariance matrix that solves the Lyapunov equation $\mat{A}\cdot\mat{V}+\mat{V}\cdot\mat{A}^\tp+\mat{D}=0$ with the input-noise matrix
\begin{equation}
\mat{D}\,\delta(t{-}t^\prime){=}\tfrac{1}{2}\bigl[\langle\hat{\vec{R}}_\mathrm{in}(t){\otimes}\hat{\vec{R}}_\mathrm{in}(t^\prime)\rangle{+}\langle\hat{\vec{R}}_\mathrm{in}(t){\otimes}\hat{\vec{R}}_\mathrm{in}(t^\prime)\rangle^\tp\bigr]\,.
\end{equation}
\begin{figure*}[tb]
 \centering{
 \subfigure[\ $Q_\mathrm{m}=5970$, $P_\mathrm{in}=10^{-8.75}$\,W]{
 \includegraphics[width=0.4\figurewidth]{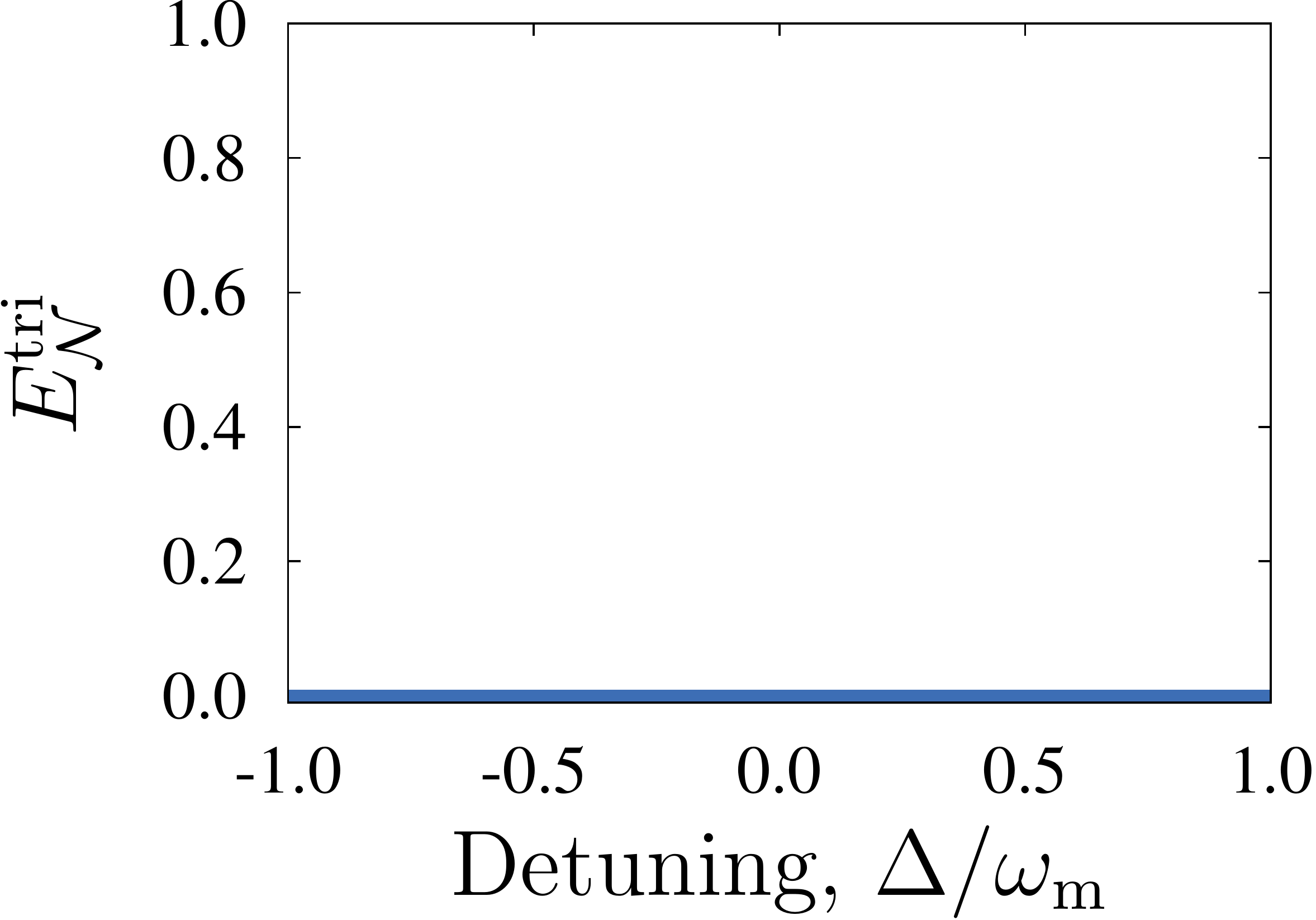}
 }\ 
 \subfigure[\ $Q_\mathrm{m}=5970$, $P_\mathrm{in}=10^{-6.0}$\,W]{
 \includegraphics[width=0.4\figurewidth]{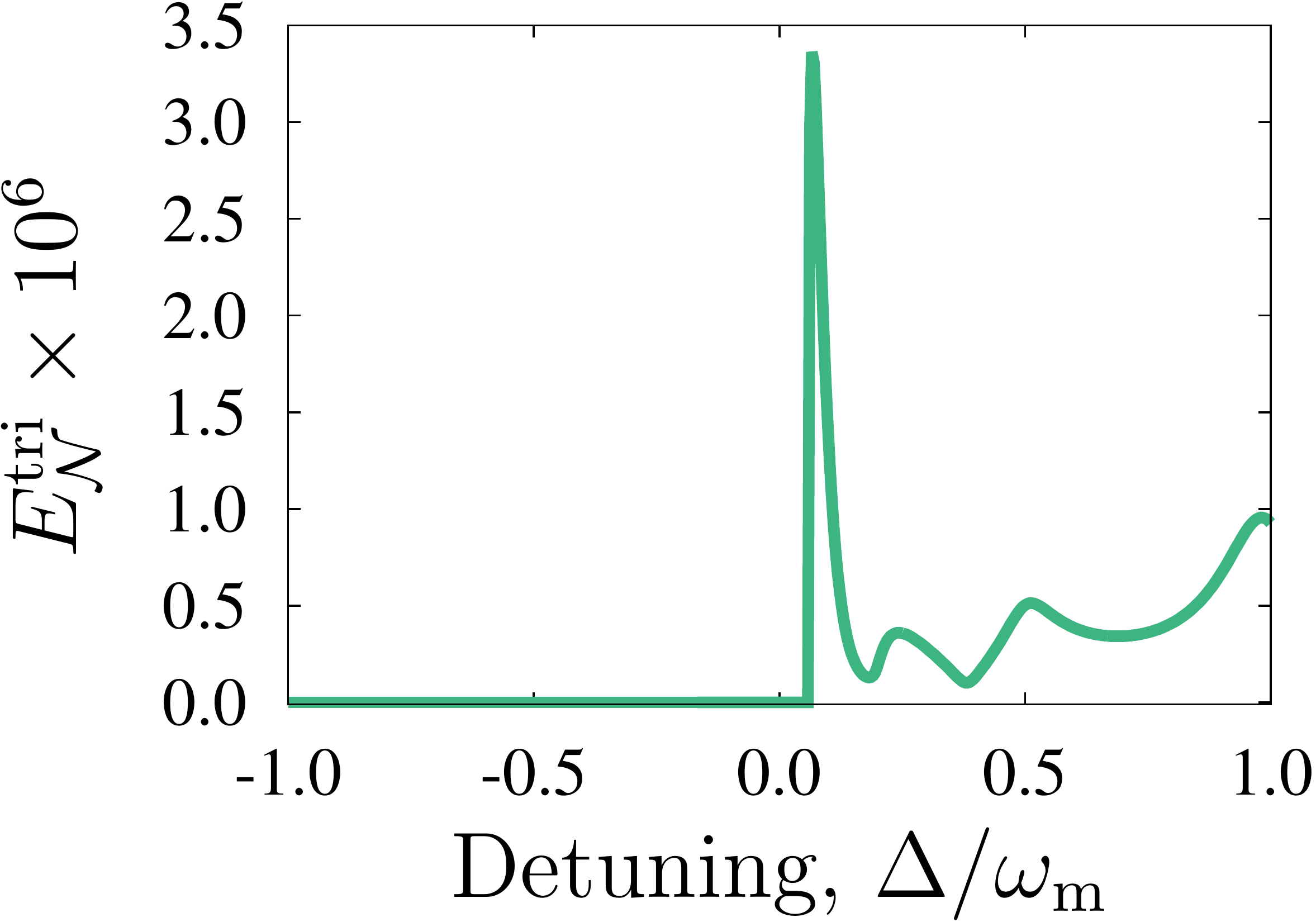}
 }\ 
 \subfigure[\ $Q_\mathrm{m}=5970$, $P_\mathrm{in}=10^{-4.5}$\,W]{
 \includegraphics[width=0.4\figurewidth]{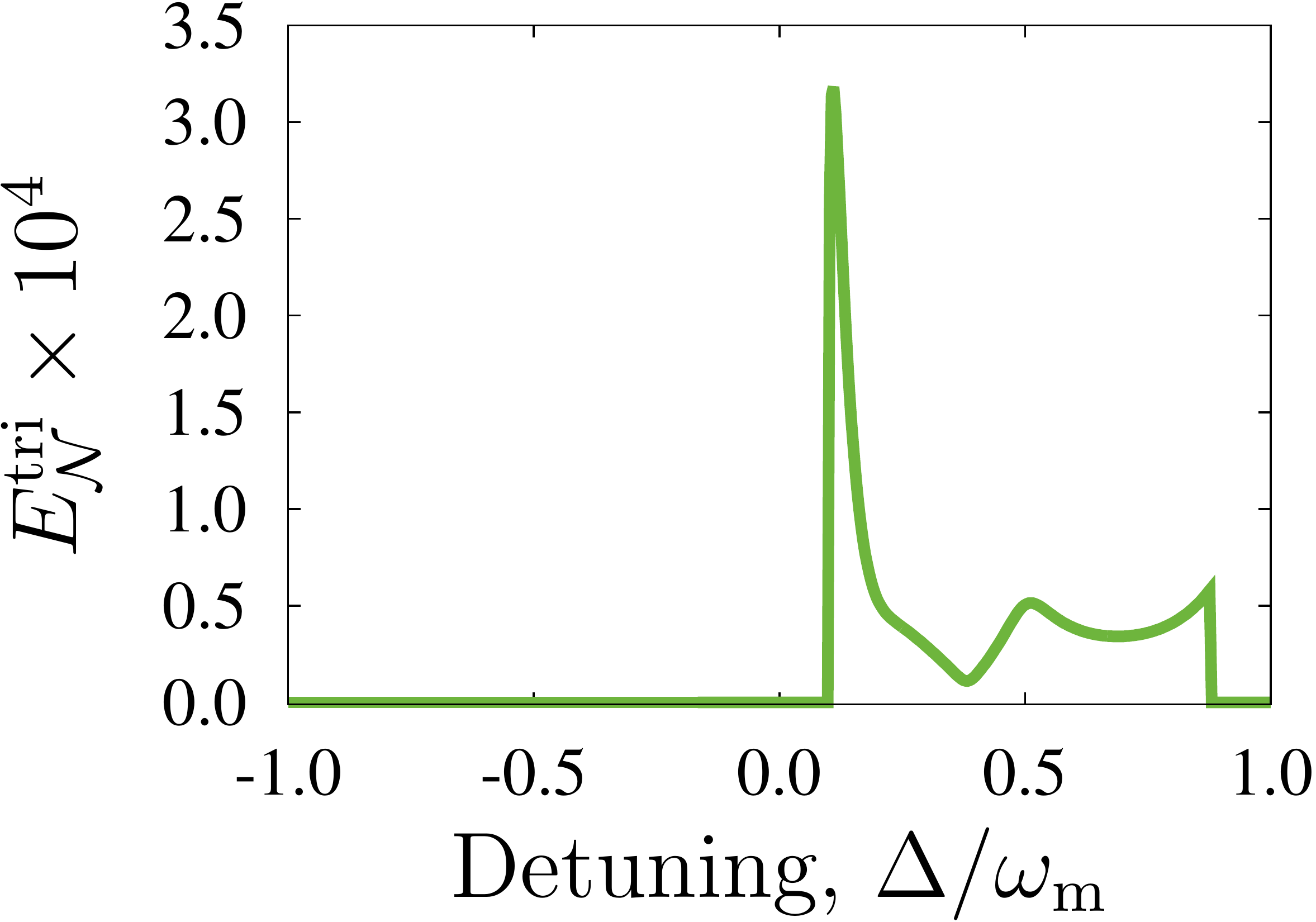}
 }\ 
 \subfigure[\ $Q_\mathrm{m}=5970$, $P_\mathrm{in}=10^{-2.5}$\,W]{
 \includegraphics[width=0.4\figurewidth]{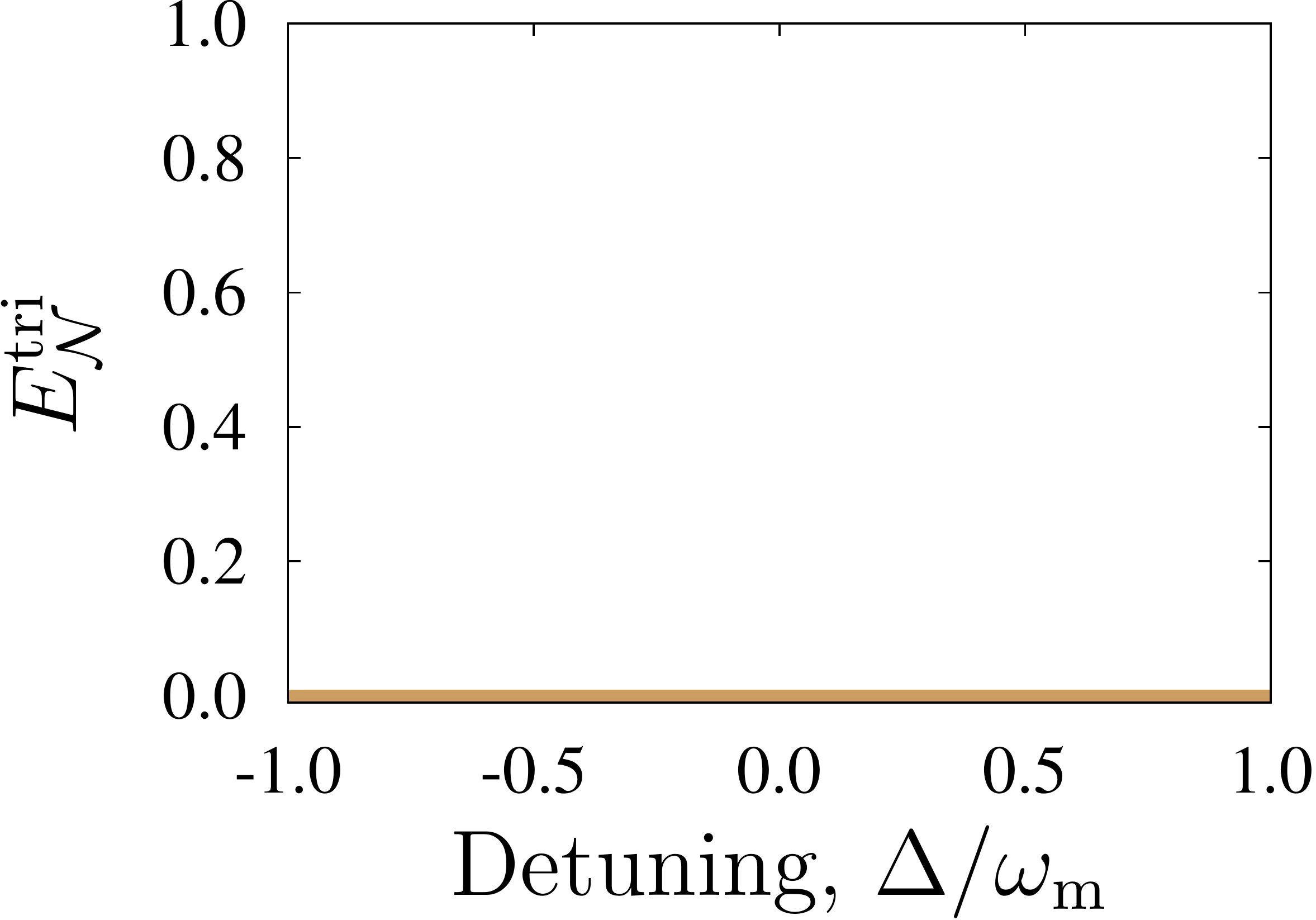}
 }\ 
 \subfigure[\ $Q_\mathrm{m}=5970$, $P_\mathrm{in}=10^{-1.5}$\,W]{
 \includegraphics[width=0.4\figurewidth]{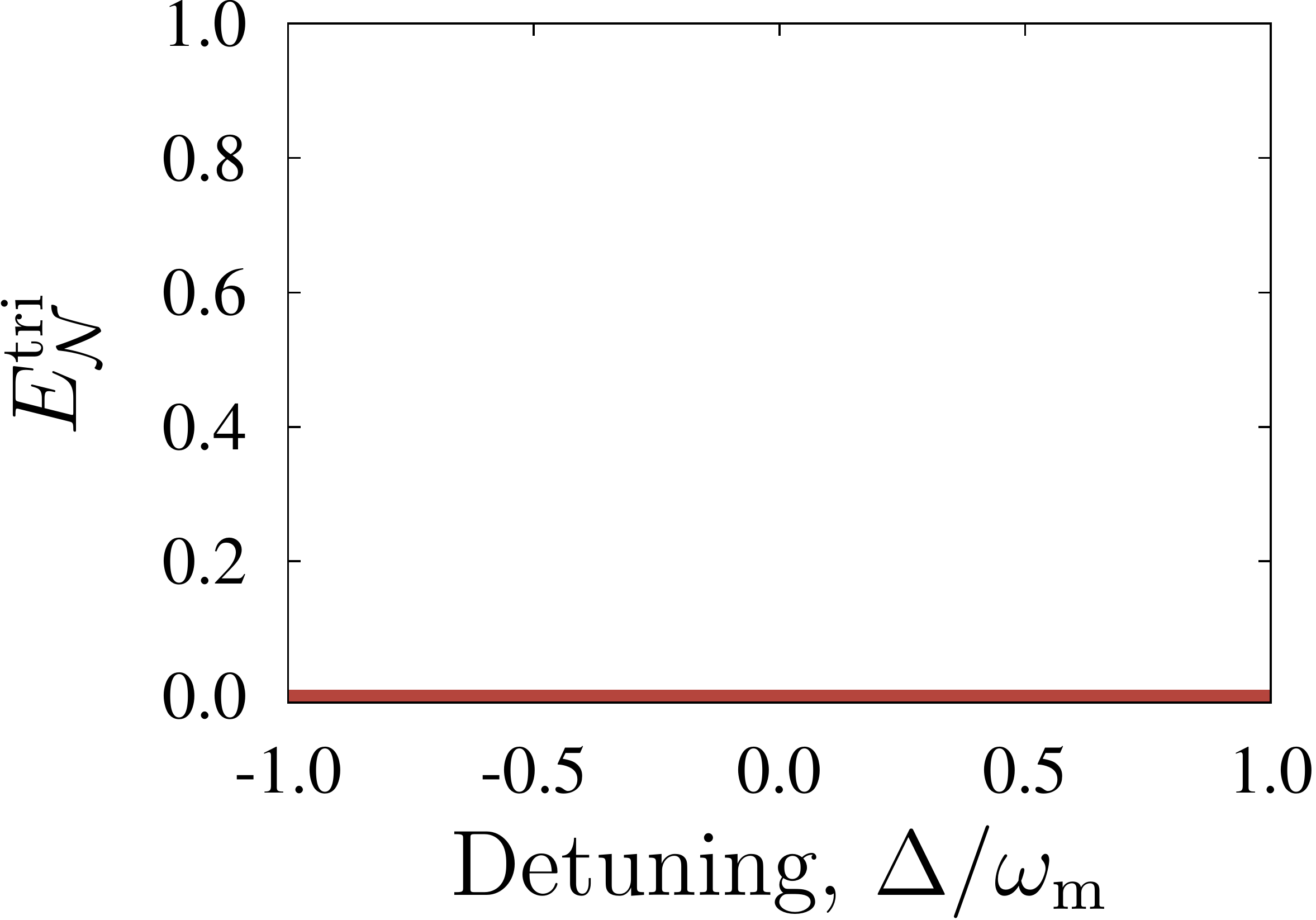}
 }\\
 \subfigure[\ $Q_\mathrm{m}=597000$, $P_\mathrm{in}=10^{-8.5}$\,W]{
 \includegraphics[width=0.4\figurewidth]{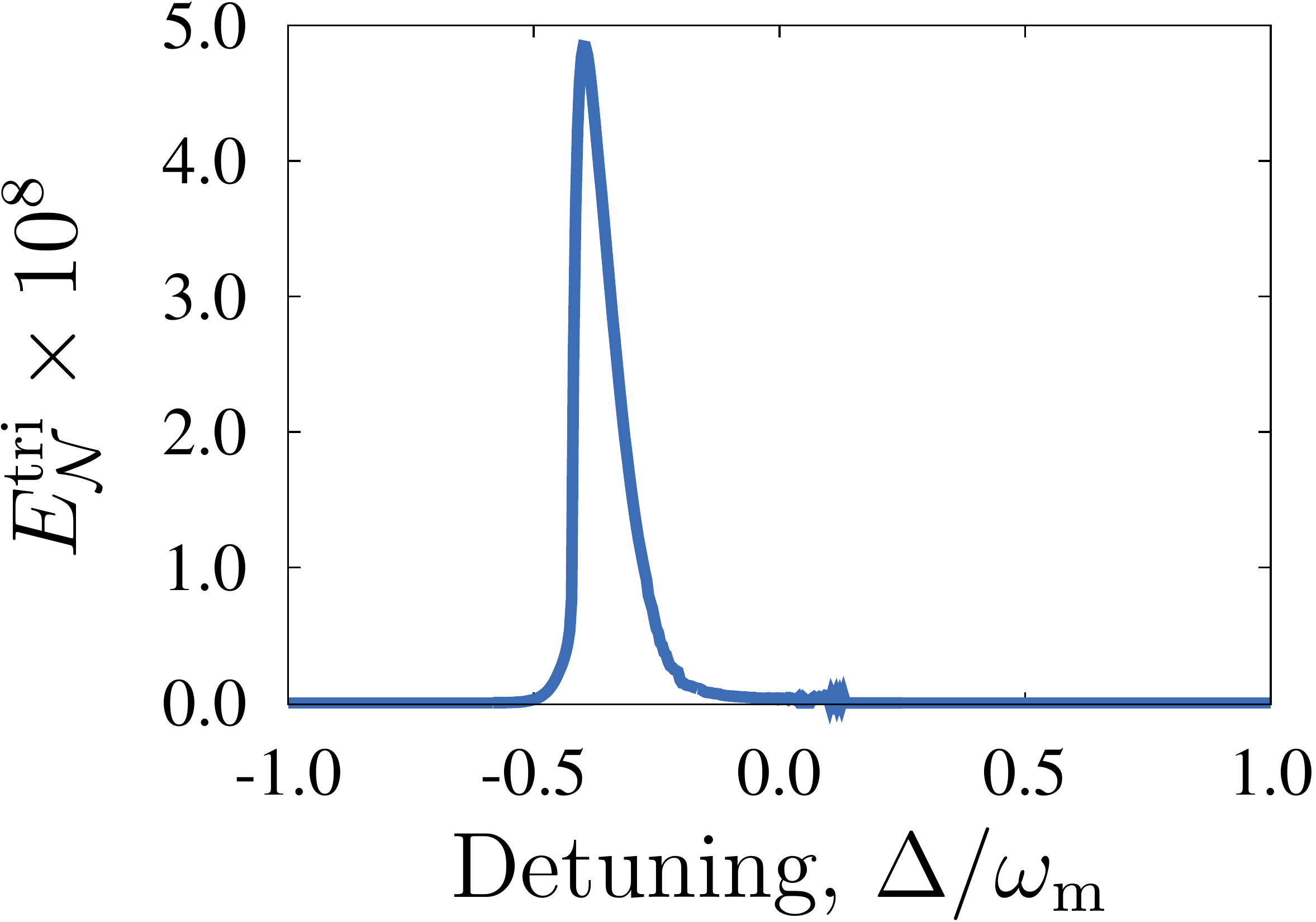}
 }\ 
 \subfigure[\ $Q_\mathrm{m}=597000$, $P_\mathrm{in}=10^{-7.5}$\,W]{
 \includegraphics[width=0.4\figurewidth]{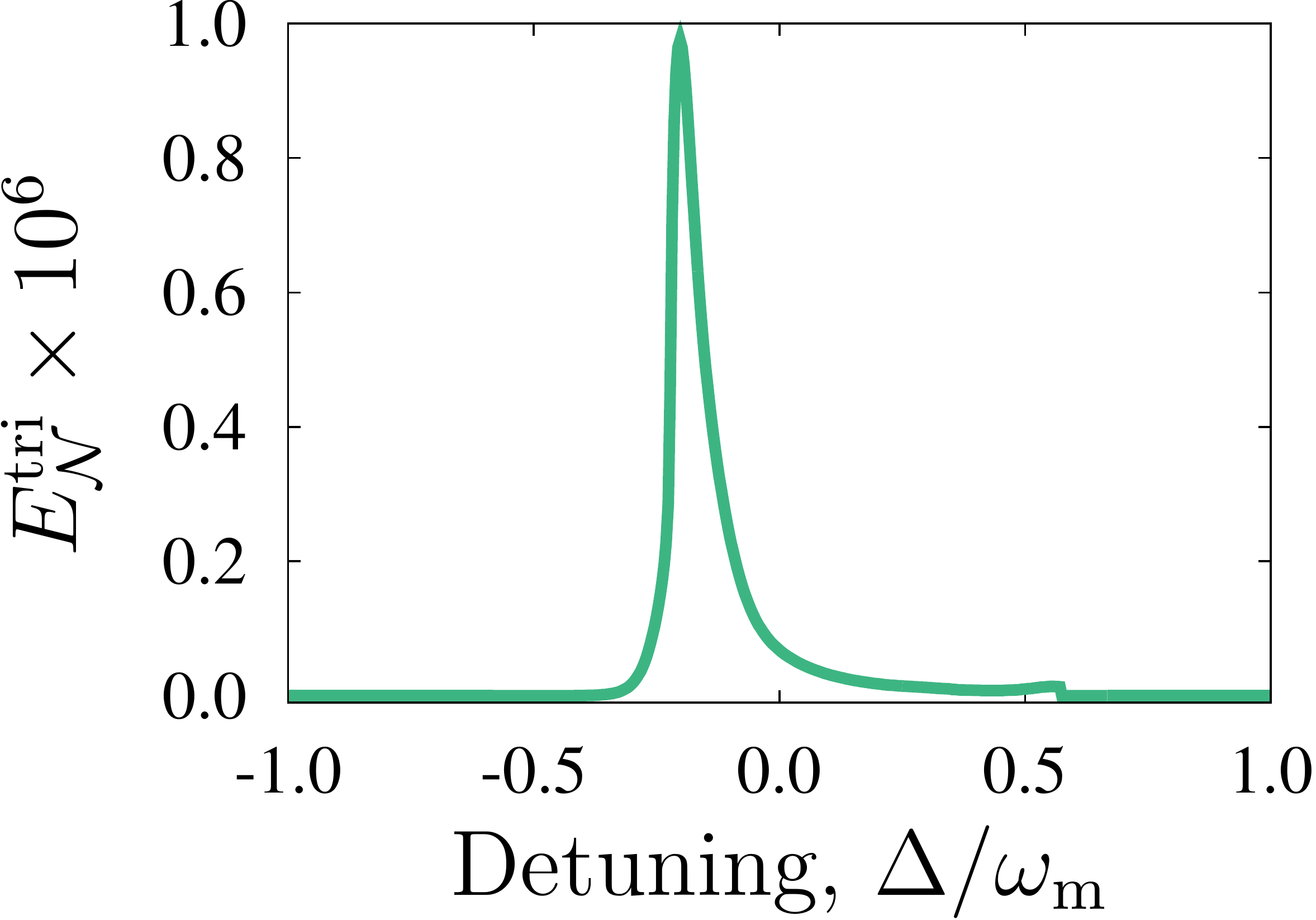}
 }\ 
 \subfigure[\ $Q_\mathrm{m}=597000$, $P_\mathrm{in}=10^{-6.5}$\,W]{
 \includegraphics[width=0.4\figurewidth]{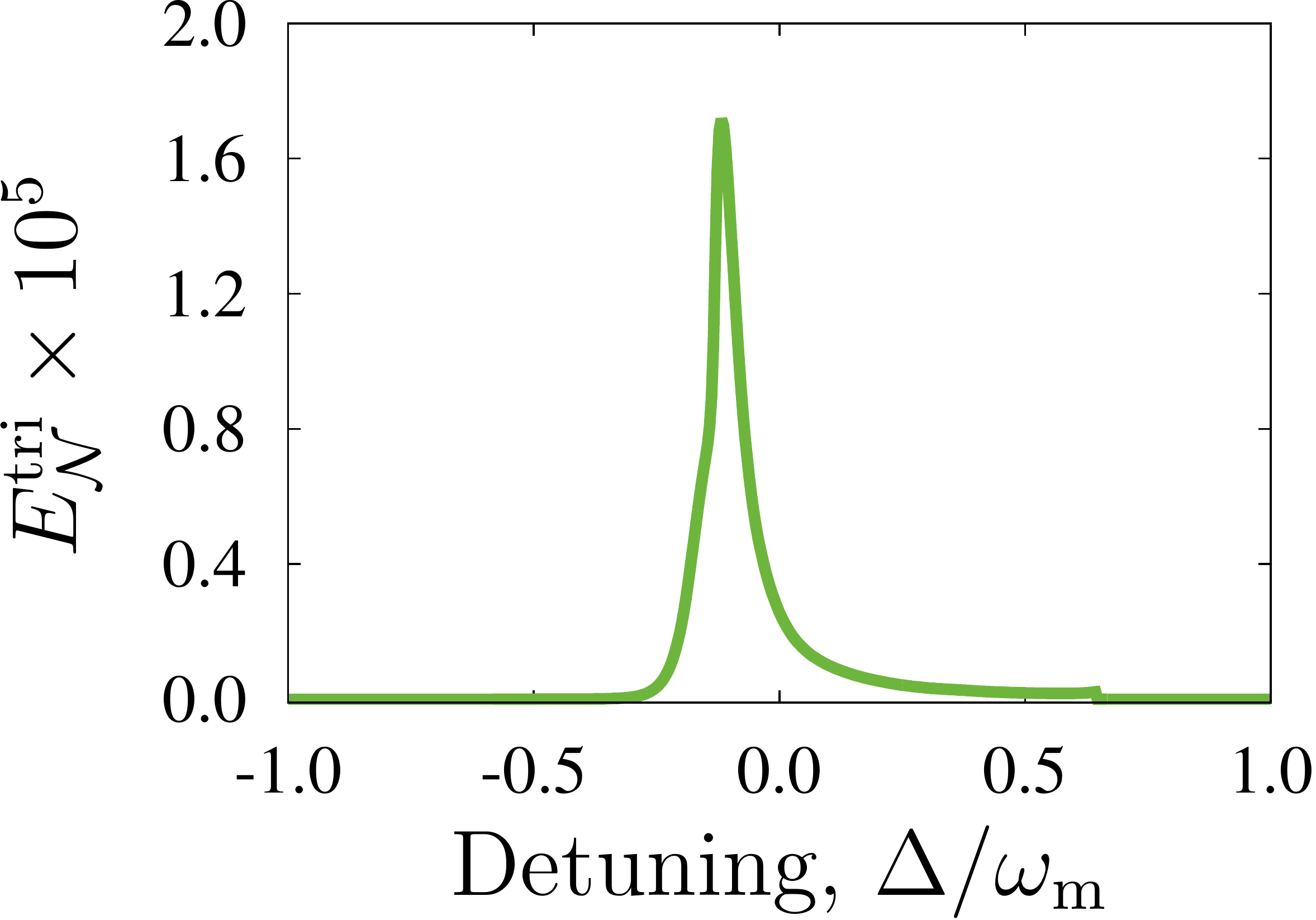}
 }\ 
 \subfigure[\ $Q_\mathrm{m}=597000$, $P_\mathrm{in}=10^{-4.0}$\,W]{
 \includegraphics[width=0.4\figurewidth]{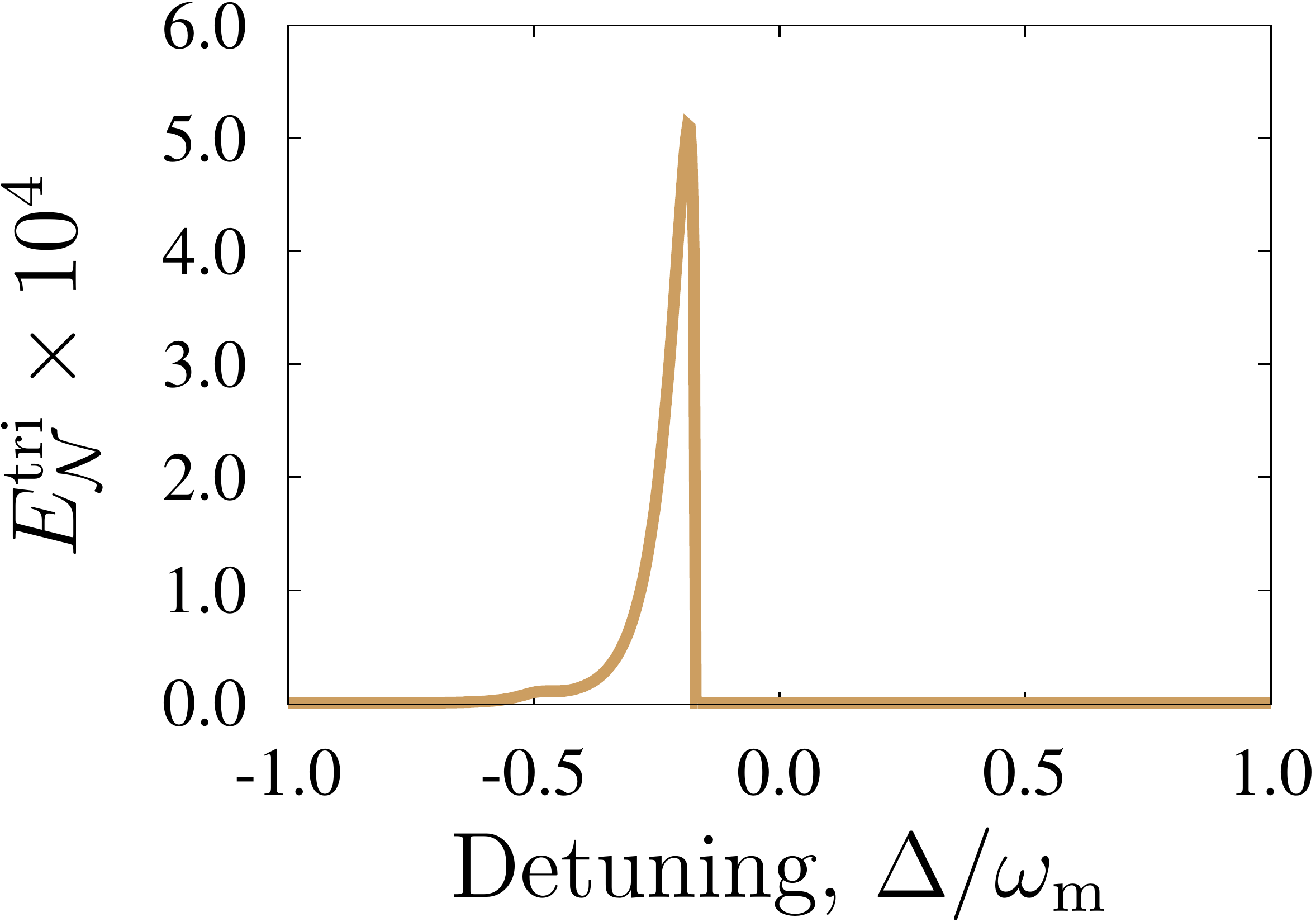}
 }\ 
 \subfigure[\ $Q_\mathrm{m}=597000$, $P_\mathrm{in}=10^{-2.5}$\,W]{
 \includegraphics[width=0.4\figurewidth]{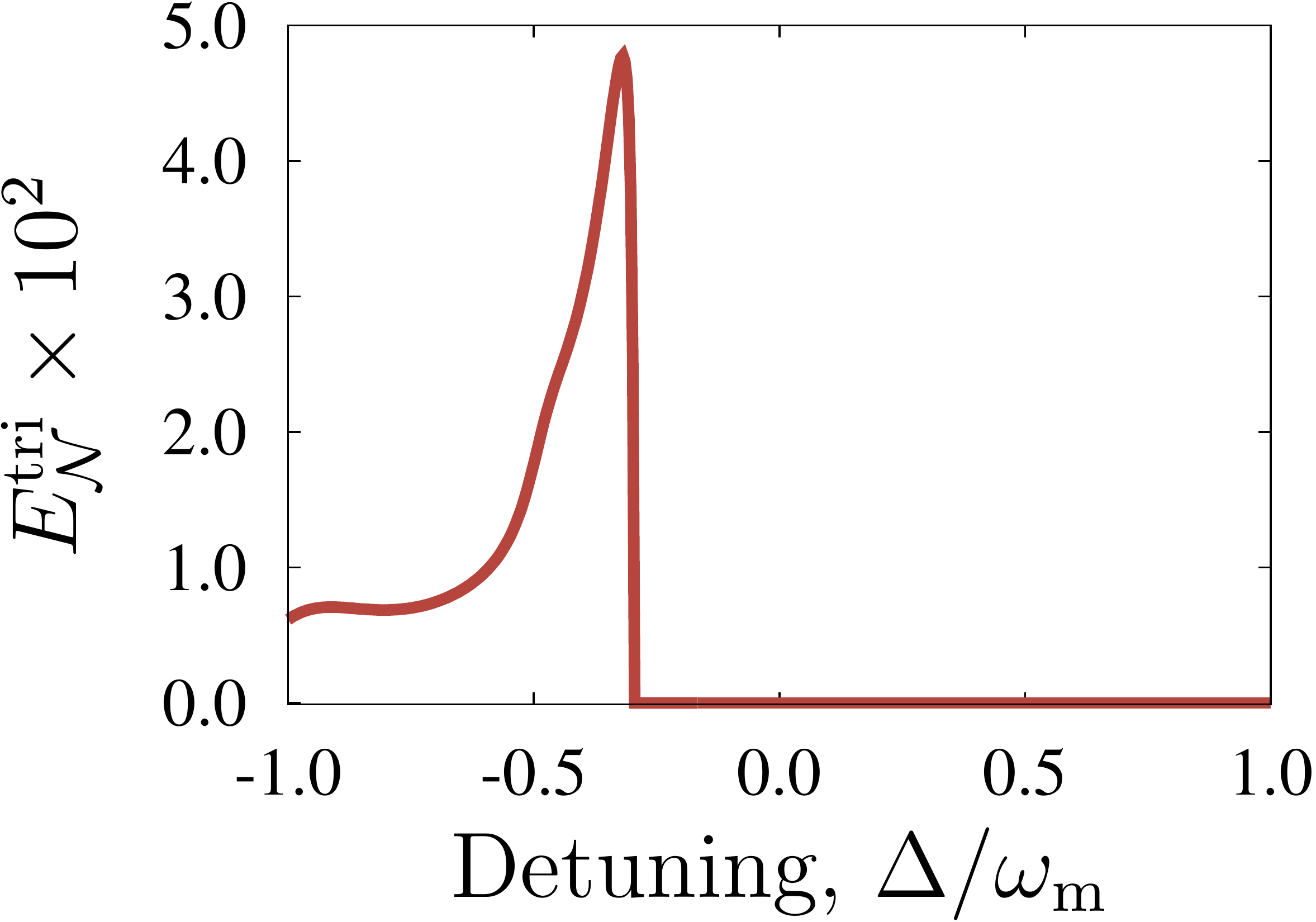}
 }
 }
 \caption{(Color online) Tripartite logarithmic negativity, $E_\mathcal{N}^\mathrm{tri}$, for five different input powers and both low- and high-$Q_\mathrm{m}$ cases.}
 \label{fig:Tripartite}
\end{figure*}
The stability condition for $\mat{A}$ can be rephrased more formally in terms of the Routh-Hurwitz criterion~\cite{Horn2008}, which we have used in our quantitative characterization of the dynamics. We note in passing that the Lyapunov equation above has a concise analytic solution for $\mat{V}$, as reported in Ref.~\cite{Horn2008}. The covariance matrix encompasses the full information on the system at hand. Here, we shall be interested in the entanglement-sharing properties of the three modes. In order to demonstrate the occurrence of genuine multipartite entanglement, we rely on the criterion based on negativity of partial transpose (NPT)~\cite{Peres1996,Horodecki1996,Simon2000} and we will make use of the logarithmic negativity as an entanglement quantifier~\cite{Vidal2002}. For a bipartition consisting of subsystems ${\mathrm{A}}$ and ${\mathrm{B}}$ ($\mathrm{A},\mathrm{B}=\mathrm{F},\mathrm{S},\mathrm{M}$), this is defined as $E_{\cal N}^{\mathrm{A}|\mathrm{B}}=\max\bigl[0,-\ln(2\sum_k\tilde\nu_{-,k})\bigr]$, where $\{\tilde{\nu}_{-,k}\}$ is the set of symplectic eigenvalues of the covariance matrix associated with the partially transposed states of the system such that $|\tilde\nu_{-,k}|<1/2$~\cite{Adesso2004}. If either ${\mathrm{A}}$ or ${\mathrm{B}}$ are single-mode subsystems, $k=1$ regardless of the number of modes comprised in ${\mathrm{B}}$ or ${\mathrm{A}}$. Moreover, although this is not the case in general, in this situation the NPT criterion is a necessary and sufficient condition for inseparability of pure and mixed Gaussian states alike. In what follows, we characterize the entanglement structure in both reduced two-mode states and bipartite one-versus-two-mode ones. In order to do that, we use numerical values for the various constants entering the model that reflect the state-of-the-art of recent experiments. The fundamental wavelength is chosen to be $1554$\,nm~\cite{Levy2011}, at which an input power $P_\mathrm{in}=1$\,$\upmu$W corresponds to ${\overline{a}_\mathrm{in}}\approx3\times10^6$\,s$^{-1}$. Moreover, we set $\omega_\mathrm{m}=2\pi\times70\,\text{MHz},\kappa_\mathrm{m}=2\pi\times5.9\,\text{kHz}\ \text{(}Q_\mathrm{m}=5970\text{)},\kappa_\mathrm{F}=\kappa_\mathrm{S}=2\pi\times7\,\text{MHz}, g_\mathrm{F}=g_{\mathrm{S}}/2=2\pi\times1.2\,\text{kHz}, T_\mathrm{env}=0.8\,\text{K}$~\cite{Riviere2011}, and $\chi=700$\,Hz, which is within a factor of $2$ of what has been observed in Ref.~\cite{Levy2011}.
\begin{table}[b]
\caption{\label{tab:EN}Calculated logarithmic negativities for $\Delta=-\omega_\mathrm{m}$ and $P_\mathrm{in}=0.27$\,W in \frefs{fig:EntangledRegions}(c) and~(d). The rest of the parameters as in the body of the paper, with $Q_\mathrm{m}=597000$.}
\begin{tabular}{D{.}{|}{-1}|c}
\hline
\hline
\multicolumn{1}{c|}{Reduction $a|b$}&\rule[-0.4em]{0cm}{1.5em}$E^{a|b}_{\cal N}$\\
\hline
\hline
\mathrm{S}.\mathrm{M}&0.10\\
\mathrm{F}.\mathrm{M}&0.42\\
\mathrm{F}.\mathrm{S}&0.01\\
\hline
\hline
\end{tabular}\hskip1cm
\begin{tabular}{D{.}{|}{-1}|c}
\hline
\hline
\multicolumn{1}{c|}{Bipartition $a|bc$}&\rule[-0.4em]{0cm}{1.5em}$E^{a|bc}_{\cal N}$\\
\hline
\hline
\mathrm{F}.\mathrm{SM}&0.44\\
\mathrm{S}.\mathrm{FM}&0.15\\
\mathrm{M}.\mathrm{FS}&0.45\\
\hline
\hline
\end{tabular}
\end{table}
\par
For these choices, \fref{fig:EntangledRegions} summarizes both the entanglement in one-versus-one-mode reduced states and one-versus-two-mode situations. Entanglement is analyzed in the $\Delta$--$P_\mathrm{in}$ parameter space, $P_\mathrm{in}$ being the input power. While all-optical entanglement (\ie, the entanglement within the reduction involving only the S and F subsystems) exists in a narrow strip around $\Delta\simeq0$ and is maximum on resonance, the mechanical mode is entangled more strongly with S near $\Delta=\pm~\omega_\mathrm{m}$, and with F close to $\Delta=\pm~\omega_\mathrm{m}/2$. At low (yet still quite sizeable) values of the mechanical quality factor, the region corresponding to $\Delta<0$ is largely associated with separability of any two-mode reduction, except the narrow strip at $\Delta\simeq0$ mentioned above, which witnesses the fact that, in these conditions, the direct nonlinear coupling between the optical modes overcomes any entangling power of the optomechanical mechanism. Larger values of $Q_{\mathrm{m}}$, on the other hand, give rise to non-negligible areas of (even strong) optomechanical entanglement involving both the M--S pair and the M--F one. This is clearly shown in \fref{fig:EntangledRegions}(c) where, remarkably, we find that even the all-optical entanglement is affected, spreading quite considerably in regions where, at lower $Q_{\mathrm{m}}$, we had $E^{\mathrm{SF}}_{\cal N}=0$. This result might be interpreted as arising from the indirect coupling between the two optical modes, whose interaction is ruled not only by their direct nonlinear coupling but also by a detuning-dependent effective one mediated (quasi-coherently) by the mechanical mode.
\par
Also quite interesting is the behavior of the one-vs-two-mode entanglement. An investigation on these configurations is relevant in order to characterize the multipartite entanglement being possibly shared by the three subsystems. Indeed, based on the classification provided by Giedke {\it et al.}~\cite{Giedke2001b}, the simultaneous inseparability of the three possible one-versus-two-mode bipartitions in a three-mode system implies the existence of genuine tripartite entanglement. Likewise, the state is $k$-mode biseparable if there are $k$ one-versus-two-mode bipartitions with respect to which the state of the system is separable. \fref{fig:EntangledRegions}(b) and \fref{fig:EntangledRegions}(d) show the rich structure of entanglement sharing that is exhibited by our model.
\begin{figure*}[t]
 \centering
 \subfigure[\ Reductions]{
 \includegraphics[width=.5\figurewidth]{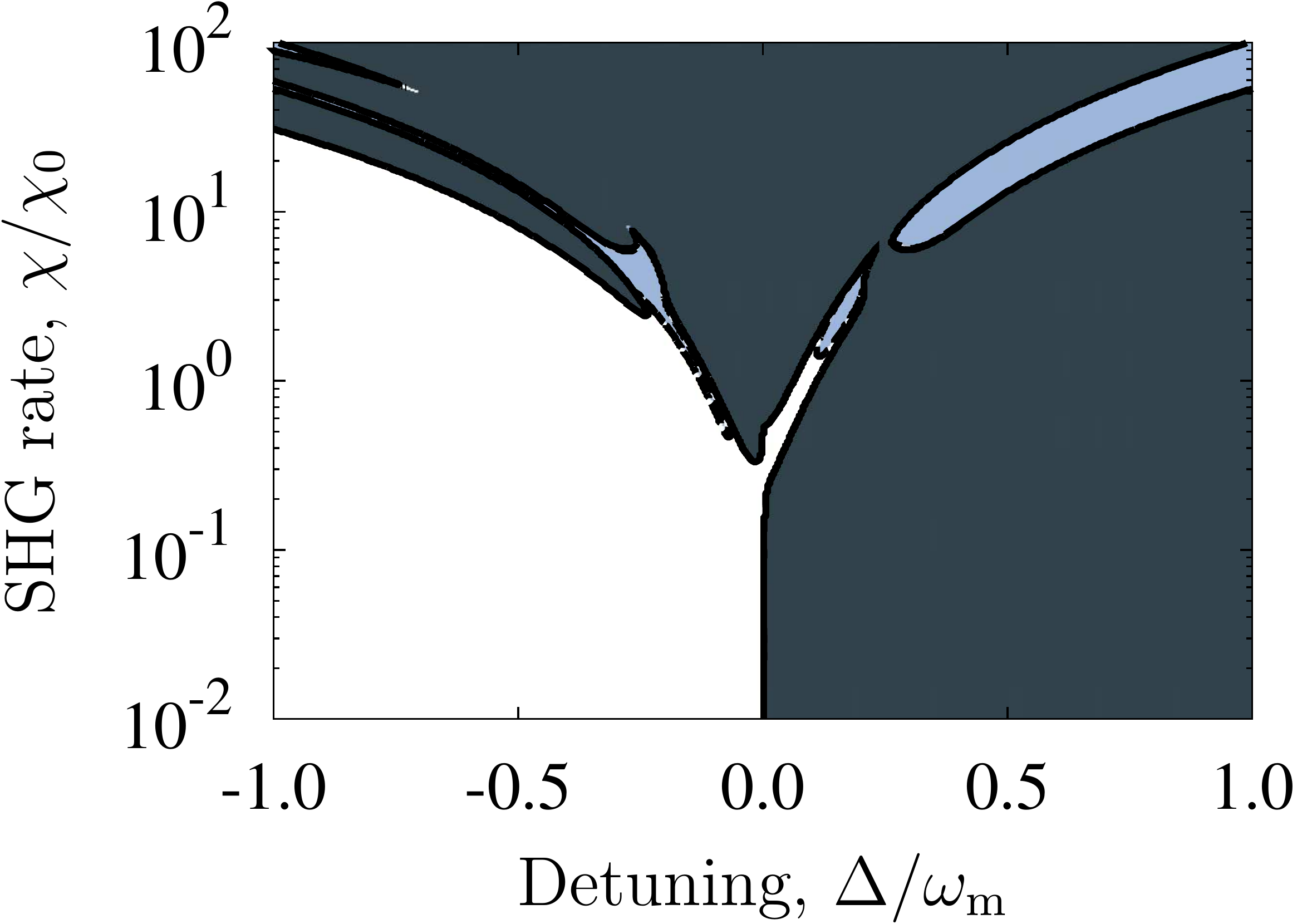}
 }
 \subfigure[\ Bipartitions]{
 \includegraphics[width=.5\figurewidth]{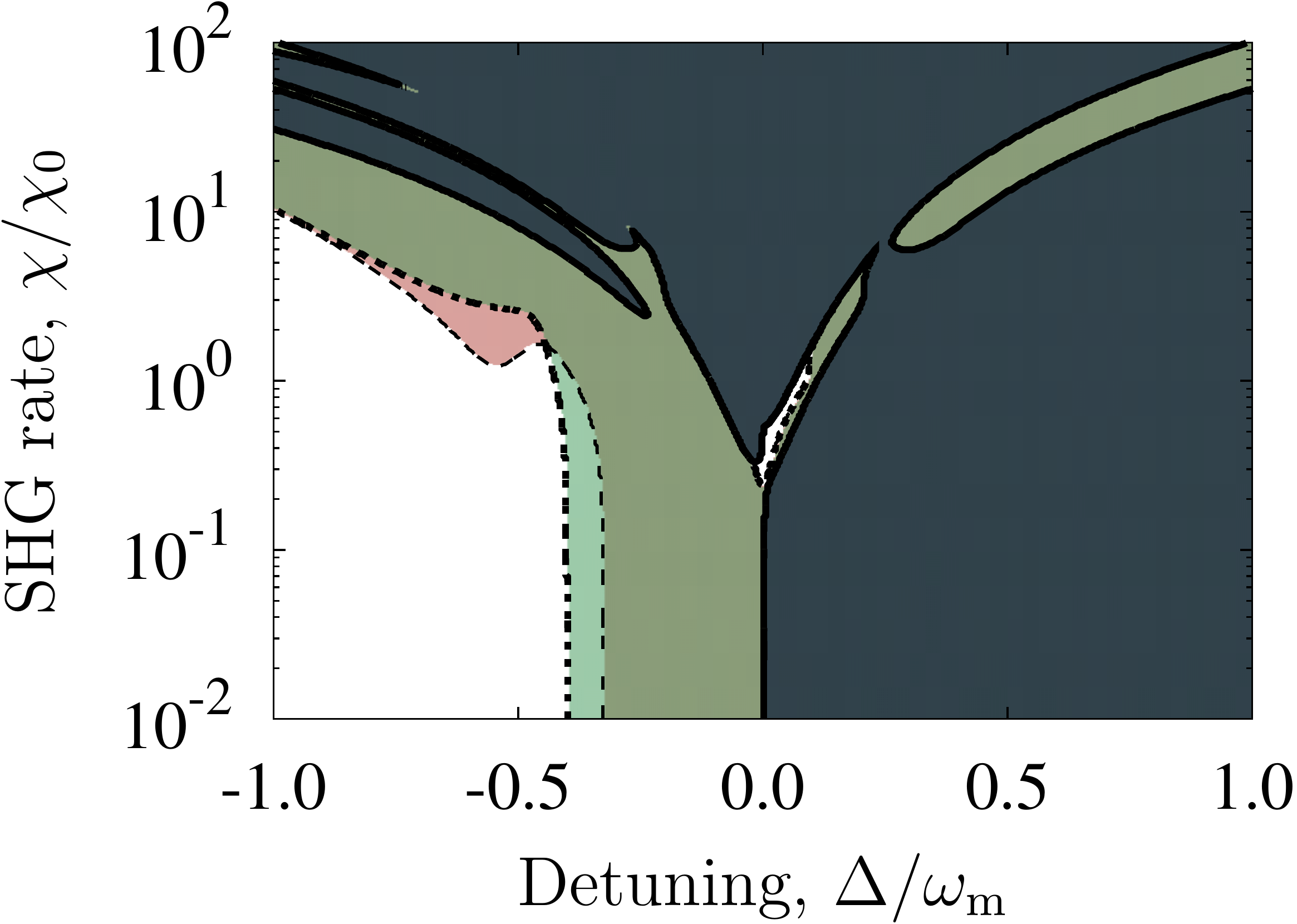}
 }
 \subfigure[\ Reductions]{
 \includegraphics[width=.5\figurewidth]{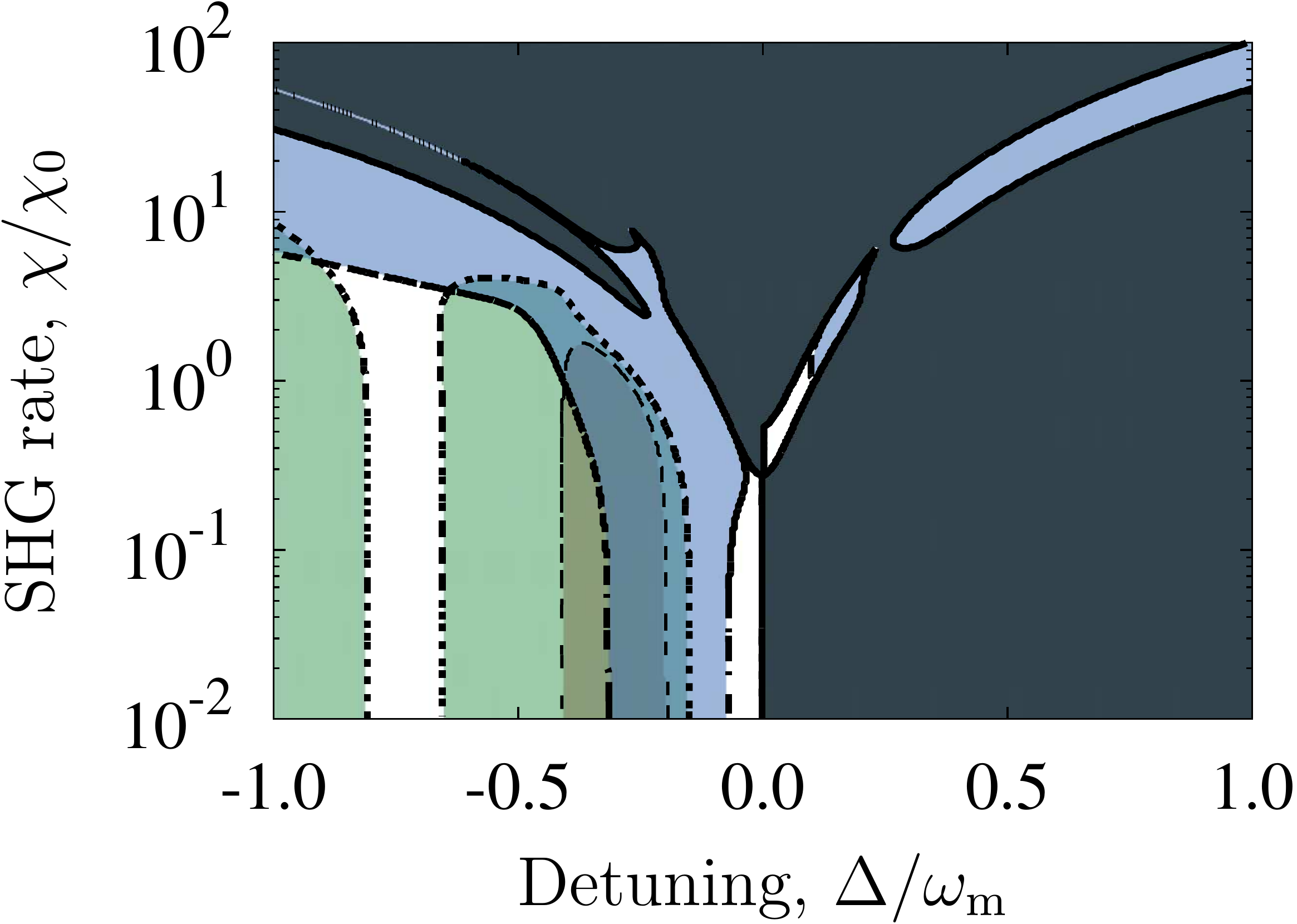}
 }
 \subfigure[\ Bipartitions]{
 \includegraphics[width=.5\figurewidth]{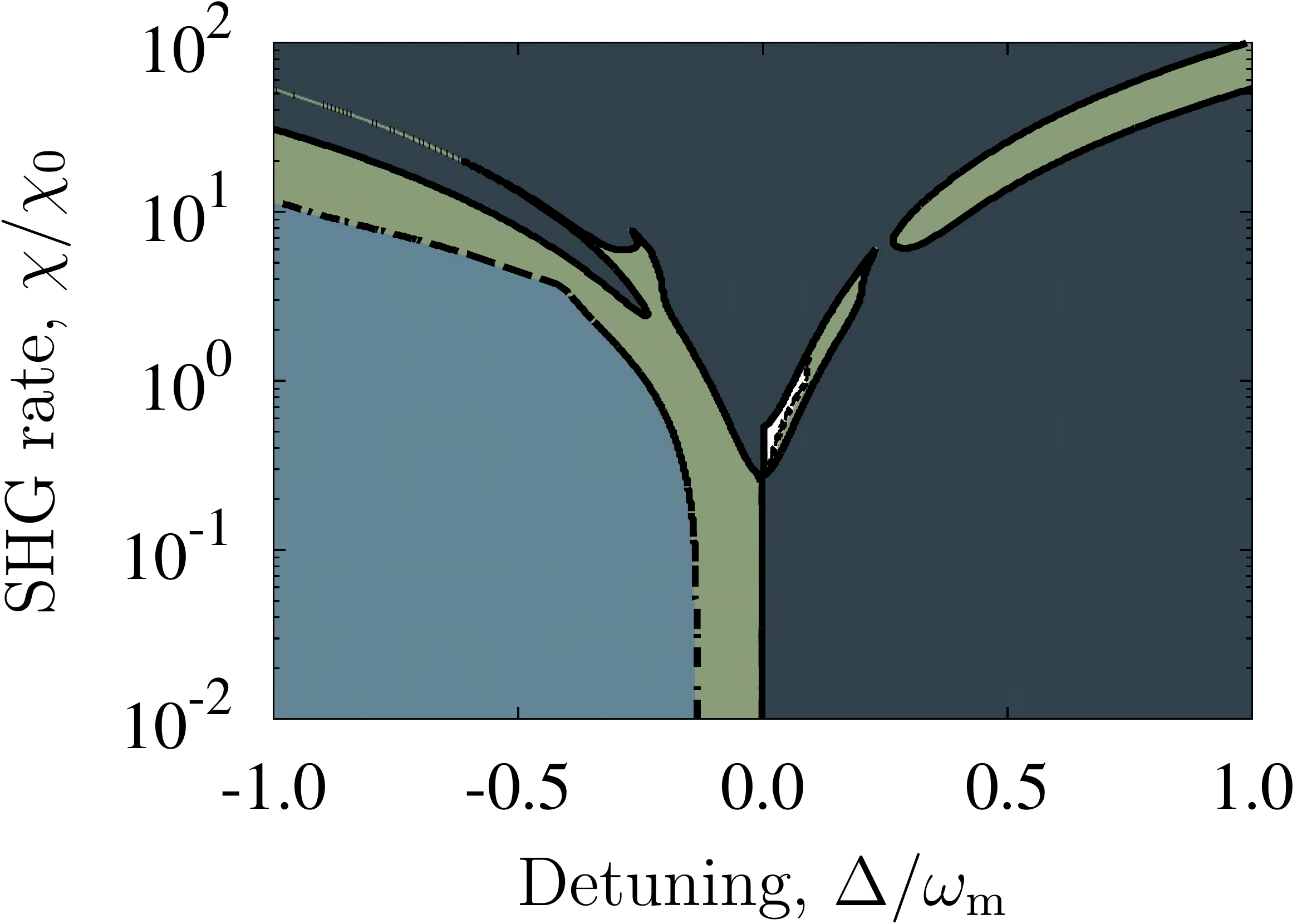}
 }
 \caption{(Color online) Similar to \fref{fig:EntangledRegions}, but varying the second-harmonic generation rate on the vertical axis. (a)~and~(b) have $Q_\mathrm{m}=5970$, whereas (c)~and~(d) have $Q_\mathrm{m}=597000$. ($\chi_0=2\pi\times700$\,Hz, $P_\mathrm{in}=10^{-3}$\,W.)}
 \label{fig:EntangledRegionsChi}
\end{figure*}
\par
One-versus-two-mode entanglement turns out to be, in general, much more robust (and larger) with respect to noise affecting the system than the entanglement in any two-mode reduction, a feature that has already been shown in other optomechanics-related investigations~\cite{DeChiara2011}. In \fref{fig:EntangledRegions}(b), regions of full three-mode inseparability are shown even for a relatively low-quality mechanical oscillator. However, in this case, noise affecting the system through the mechanics is too strong to allow for much overlap between regions of three-mode inseparability. Indeed, by increasing the mechanical quality factor by a factor of $100$, giving the results in \fref{fig:EntangledRegions}(c), the overlap between regions of one-versus-two-mode entanglement increases significantly, covering virtually the whole stability area shown in the figure. In passing, we mention that we have applied a multipartite entanglement witness for continuous-variable states, \textsc{MultiWit}, that was developed in Ref.~\cite{Hyllus2006} using semi-definite optimization methods. The use of this instrument has confirmed the genuinely tripartite nature of the entanglement at hand in the regions of overlap among the three regions of one-versus-two-mode inseparability, which excludes the possibility of having generalized three-mode biseparable states. As a quantitative illustration for the high-$Q_{\mathrm{m}}$ case, in \tref{tab:EN} we give the entanglement in any reduction and bipartition that can be singled-out in our problem, taking the values of the parameters listed above and choosing $\Delta=-\omega_\mathrm{m}$ with $P_\mathrm{in}=0.27$\,W. In order to complete our assessment, we have determined the degree of genuine tripartite entanglement across interesting regions in the full-inseparability areas. As a quantitative estimator, we have used the tripartite logarithmic negativity $E_\mathcal{N}^\mathrm{tri}$, which is a proper entanglement monotone~\cite{Sabin2008}. The results of this study are shown in~\fref{fig:Tripartite} for both the low- and high-$Q_{\mathrm{m}}$ cases and increasing optical input powers. The qualitative differences in the behavior of the tripartite entanglement is very marked at large input power:\ whilst at low mechanical quality factors the high-power tripartite entanglement is null, it extends for most of the region $\Delta\in[-\omega_{\mathrm{m}},0]$ at high mechanical quality factor, therefore leaving us with much room for maneuvering in the space of entangled three-mode states.\\
Let us finally explore the competition between the SHG process and the optomechanics in our model. As $\chi$ grows in \fref{fig:EntangledRegionsChi}(a) and \fref{fig:EntangledRegionsChi}(c), the entanglement for the F$|$M and S$|$M reductions decreases, whereas that for F$|$S covers an ever-larger area of the parameter space. The behavior of the entanglement in the bipartitions, \fref{fig:EntangledRegionsChi}(b) and \fref{fig:EntangledRegionsChi}(d), is similar and also easily understood on an intuitive basis:\ a bigger $\chi$ leads to larger regions of entanglement for the two bipartitions that involve one of the optical mode on its own (\ie, F$|$MS and S$|$MF), but a contraction in the parameter space where entanglement in the M$|$FS bipartition is observed.

\section{Inferring the state of the system}
\begin{figure}[t]
 \includegraphics[width=\linewidth]{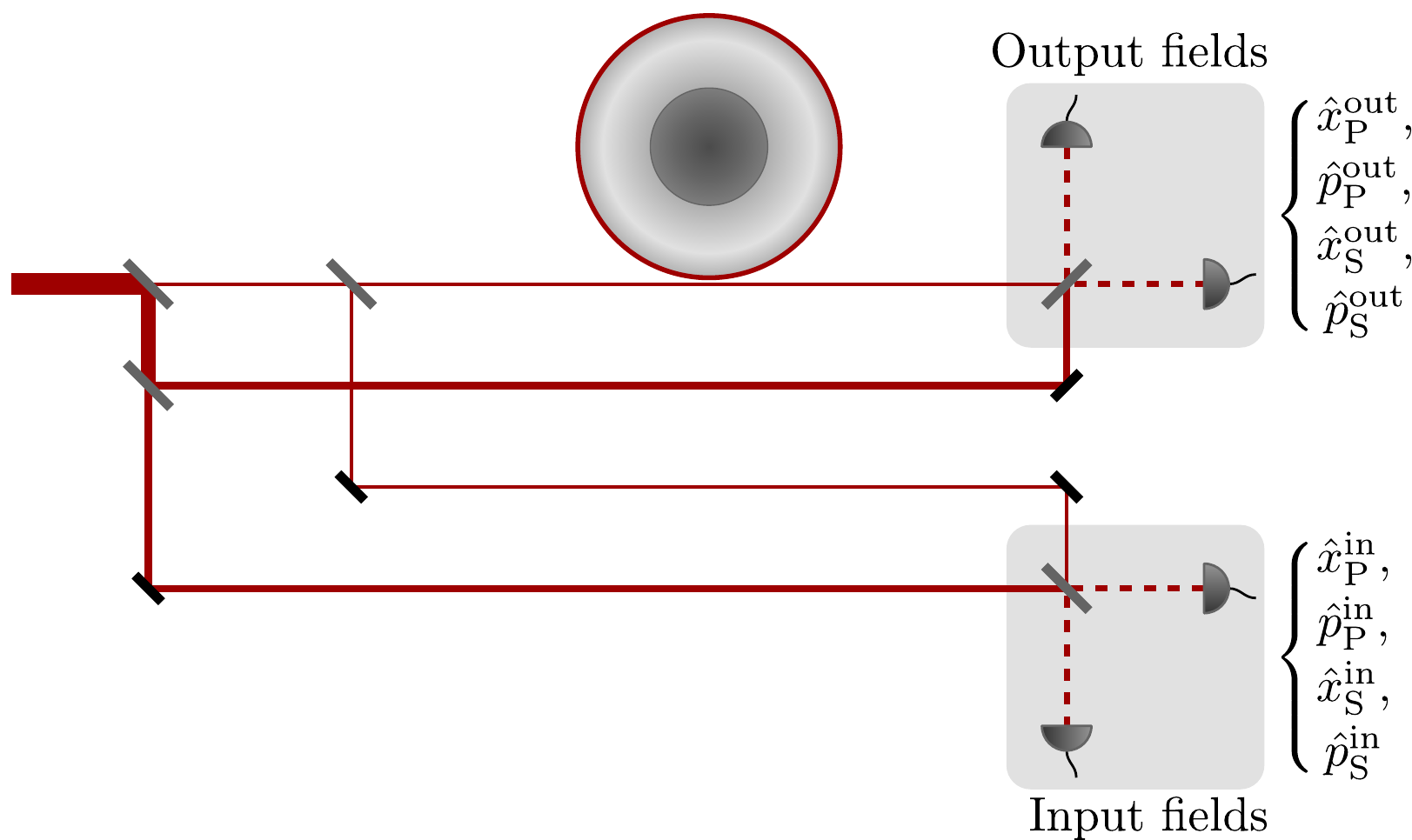}
 \caption{Schematic of detection system. Part of the input field is used as the local oscillator of two homodyne detectors (this necessitates adding the second harmonic onto the local oscillator), one recording the field quadratures before the toroid, and one after.}
 \label{fig:Detection}
\end{figure}

The inference of the full state of an optomechanical system is a major practical challenge~\cite{Paternostro2007}, mainly due to the fact that the mechanical quadratures are not directly accessible to an experiment. Here we propose a technique, which requires the use of the system drawn schematically in \fref{fig:Detection}, that allows us to infer the mechanical quadratures indirectly. Initially, we assume that the homodyne detectors needed in the scheme have infinite bandwidth; we shall account for the finite bandwidth of any realistic apparatus later on. Given such a system, one has access to the four input (two each for the fundamental and the second harmonic) and four output quadratures. One can then infer the intra-cavity optical modes using the input-output relation
\begin{equation}
\hat{x}_\mathrm{F}=\bigl(\hat{x}_\mathrm{F}^\mathrm{out}-\hat{x}_\mathrm{F}^\mathrm{in}\bigr)/{\sqrt{2\kappa_\mathrm{F}}},
\end{equation}
and similarly for the rest of the quadratures. It is clear that very good characterization of the system, including knowledge of all the coupling constants and the effect of vacuum input noise~\cite{Paternostro2007} is required to infer the intra-cavity quadratures accurately. Once the four optical intra-cavity quadratures are known, it is natural to ask how the mechanical quadratures can be inferred. Indeed, the key to our proposal is noticing that the inferred quadratures are obtained as a time trace. One can therefore make direct use of the equations of motion to obtain a time trace for the mechanical quadrature $\hat{x}$ and $\hat{p}$. The covariances of these inferred quadratures can finally be used to build an inferred covariance matrix, $\tilde{\mat{V}}$. In the limit of infinite detector bandwidth, $\tilde{\mat{V}}=\mat{V}$.

We use a simple model for including the effects of a finite detector bandwidth. Let the bandwidth of the detector be $\tau$. Then the point-spread function of the detector is taken to be
\begin{equation}
f(t){=}\frac{\Theta(t)-\Theta(t-\tau)}{\tau}=\begin{cases}
0\quad&\textrm{for}\ t<0\ \text{or}\ t>\tau\\
\frac{1}{\tau}\quad&\textrm{for}\ t\in(0,\tau)\\
\frac{1}{2\tau}\quad&\textrm{for}\ t=0\ \text{or}\ t=\tau
\end{cases}\,,
\end{equation}
where $\Theta(t)$ is the Heaviside step function. The label $t$ is understood as the time at which the measurement was performed. Normalization requires that $\int_{-\infty}^\infty f(t)\,\rmd t=1$. For each operator $\hat{a}(t)$ we assign an inferred operator $\tilde{a}(t){\equiv}(f{\ast}\hat{a})(t)=\int_{-\infty}^\infty f(t-s)\hat{a}(s)\,\rmd s$ as the convolution of $f(t)$ with $\hat{a}(t)$. It can easily be shown that $\frac{\rmd}{\rmd t}(f\ast\hat{a})(t)=(f\ast\dot{\hat{a}})(t)$. This means that we can infer the value of any $\dot{\hat{a}}(t)$ by calculating the time derivative of the \emph{inferred} $\tilde{a}(t)$.\par
With this at hand, we can finally show that the inferred covariance matrix, at the steady-state, is given by
\begin{equation}
\begin{aligned}
\tilde{\mat{V}}&=\bigl(\mat{A}\tau\bigr)^{-1}\left\{\bigl(e^{\mat{A}\tau}-1\bigr)\mat{V}\bigl(e^{\mat{A}\tau}-1\bigr)^\tp\right.\\
&+\int_0^{\tau}\left.\bigl[e^{\mat{A}(\tau-s)}-1\bigr]\mat{D}\bigl[e^{\mat{A}(\tau-s)}-1\bigr]^\tp\,\rmd s\right\}\bigl(\mat{A}\tau\bigr)^{-\tp}.
\end{aligned}
\end{equation}
Modern homodyne detectors can operate with a bandwidth of the order of $10$\,GHz, which is much larger than typical values of $\omega_\mathrm{m}$ in the micro-mechanical domain. Therefore, it is understood that $\tau$ is by far the shortest timescale of the system. It then suffices to expand $\tilde{\mat{V}}$ to first order in $\tau$, which gives
\begin{equation}
\tilde{\mat{V}}=\mat{V}-\tfrac{1}{6}\tau\mat{D}\,.
\end{equation}
These expressions hint at the tantalizing possibility that by increasing $\tau$ electronically one could deduce the value of $\tilde{\mat{V}}$ for \emph{vanishing} $\tau$, and therefore infer $\mat{V}$ itself. In our numerical exploration, the fidelity~\cite{Olivares2006} for the inference of the mechanical mode using this method was above $99$\% when $\tau$ corresponded to a bandwidth of $500$\,MHz.

\section{Conclusions and outlook}
We have presented a system that combines a nonlinear optical process with optomechanics in a very natural manner. Its monolithic design makes it very attractive for experimental, or even technological, applications. Indeed, the system we presented is based on technology that is inherently compatible with integration on optoelectronic chips. Our investigation concentrated on the dynamics of the system, but we also addressed the problem of the actual detection of the intra-cavity state by outlining a method involving homodyning all the input and output quadratures to infer the covariance matrix for the three intra-cavity modes. Lastly, a numerical example using constants from recent experiments was used to illustrate the feasibility of observing these effects in a realistic system.\\
Looking further ahead, one can envisage several of these structures sharing a common photonic ``bus'' whose function is to populate the optical mode of each toroid at the fundamental frequency. The SHG process and optomechanics could then be used to create an entangled state of the mechanics with the second-harmonic field, thereby generating for each structure an optical field, which can be routed out of the photonic bus without losses, due to the large separation in frequencies, that is entangled to the mechanical mode of the toroid. We remark that this system lends itself naturally to the distribution and certification of optomechanical entanglement as per the protocol recently proposed in Ref.~\cite{Abdi2012}.

\section*{Acknowledgements}
AX acknowledges financial support from the Royal Commission for the Exhibition of 1851. MB is supported by a FASTQUAST ITN Marie Curie fellowship. MP is supported by the UK EPSRC through a Career Acceleration Fellowship and the ``New Directions for EPSRC Research Leaders'' initiative (EP/G004759/1).

\section*{Appendix}
Here we provide the explicit form of the drift matrix $\mat{A}$ for our problem, which reads
\begin{widetext}
\begin{equation}
\label{eq:DriftMatrix}
\mat{A}{=}\left[
\begin{matrix}
-\kappa_\mathrm{F}+2\chi{{\overline{a}^\mathrm{r}_\mathrm{S}}}&-\Delta+2\chi{{\overline{a}^\mathrm{i}_\mathrm{S}}}&\phantom{+}2\chi{{\overline{a}^\mathrm{r}_\mathrm{F}}}&2\chi{{\overline{a}^\mathrm{i}_\mathrm{F}}}&-\sqrt{2}\,g_\mathrm{F}{{\overline{a}^\mathrm{i}_\mathrm{F}}}&0\\
\Delta+2\chi{{\overline{a}^\mathrm{i}_\mathrm{S}}}&-\kappa_\mathrm{F}-2\chi{{\overline{a}^\mathrm{r}_\mathrm{S}}}&-2\chi{{\overline{a}^\mathrm{i}_\mathrm{F}}}&2\chi{{\overline{a}^\mathrm{r}_\mathrm{F}}}&\phantom{+}\sqrt{2}\,g_\mathrm{F}{{\overline{a}^\mathrm{r}_\mathrm{F}}}&0\\
-2\chi{{\overline{a}^\mathrm{r}_\mathrm{F}}}&\phantom{+}2\chi{{\overline{a}^\mathrm{i}_\mathrm{F}}}&-\kappa_\mathrm{S}&-2\Delta&-\sqrt{2}\,g_\mathrm{S}{{\overline{a}^\mathrm{i}_\mathrm{S}}}&0\\
-2\chi{{\overline{a}^\mathrm{i}_\mathrm{F}}}&-2\chi{{\overline{a}^\mathrm{r}_\mathrm{F}}}&2\Delta&-\kappa_\mathrm{S}&\phantom{+}\sqrt{2}\,g_\mathrm{S}{{\overline{a}^\mathrm{r}_\mathrm{S}}}&0\\
0&0&0&0&0&\omega_\mathrm{m}\\
\sqrt{2}\,g_\mathrm{F}{{\overline{a}^\mathrm{r}_\mathrm{F}}}&\sqrt{2}\,g_\mathrm{F}{{\overline{a}^\mathrm{i}_\mathrm{F}}}&\sqrt{2}\,g_\mathrm{S}{{\overline{a}^\mathrm{r}_\mathrm{S}}}&\sqrt{2}\,g_\mathrm{S}{{\overline{a}^\mathrm{i}_\mathrm{S}}}&-\omega_\mathrm{m}&-2\kappa_\mathrm{m}
\end{matrix}
\right]
\end{equation}
with $\overline{a}_{\mathrm{j}}=\overline{a}^\mathrm{r}_{\mathrm{j}}+i\overline{a}^\mathrm{i}_{\mathrm{j}}~(j=\mathrm{F,S})$. As done in the main body of the paper, we assume that ${\overline{a}_\mathrm{F}}$ is real and $g_\mathrm{S}=2g_\mathrm{F}$. This can always be done by dropping the corresponding assumption on ${\overline{a}_\mathrm{in}}$ and choosing its phase appropriately. We then define the parameters $\alpha=g_\mathrm{F}\big/(\sqrt{2}\,\chi)$ and $\beta=\chi{\overline{a}_\mathrm{F}^\mathrm{r}}$ and rewrite the drift matrix as
\begin{equation}
\mat{A}=\left[
\begin{array}{cccccc}
-\kappa_\mathrm{F}-\frac{2\kappa_\mathrm{S}}{4\Delta^2+\kappa_\mathrm{S}^2}\beta^2&-\Delta-\frac{4\Delta}{4\Delta^2+\kappa_\mathrm{S}^2}\beta^2&\phantom{+}2\beta&0&0&0\\
\Delta+\frac{4\Delta}{4\Delta^2+\kappa_\mathrm{S}^2}\beta^2&-\kappa_\mathrm{F}+\frac{2\kappa_\mathrm{S}}{4\Delta^2+\kappa_\mathrm{S}^2}\beta^2&0&2\beta&2\alpha\beta&0\\
-2\beta&0&-\kappa_\mathrm{S}&-2\Delta&\phantom{+}\frac{8\Delta}{4\Delta^2+\kappa_\mathrm{S}^2}\alpha\beta^2&0\\
0&-2\beta&2\Delta&-\kappa_\mathrm{S}&-\frac{4\kappa_\mathrm{S}}{4\Delta^2+\kappa_\mathrm{S}^2}\alpha\beta^2&0\\
0&0&0&0&0&\omega_\mathrm{m}\\
2\alpha\beta&0&-\frac{4\kappa_\mathrm{S}}{4\Delta^2+\kappa_\mathrm{S}^2}\alpha\beta^2&-\frac{8\Delta}{4\Delta^2+\kappa_\mathrm{S}^2}\alpha\beta^2&-\omega_\mathrm{m}&-2\kappa_\mathrm{m}
\end{array}
\right],
\end{equation}
\end{widetext}
showing that the steady-state of the system is a universal function of $\alpha$ and $\beta$. In particular, at a fixed value for $\beta$ the dynamics is determined solely by the value of the ratio of the coupling constants $\alpha$. 

\end{document}